\documentclass[preprintnumbers,nofootinbib,showkeys,showpacs,amsmath,amssymb]{revtex4}
\usepackage{amsmath,amssymb,graphics,epsfig,subfigure}
\usepackage{color}
\usepackage{multirow}
\usepackage{booktabs}
\usepackage{changes}
\usepackage{footnote}

\usepackage{hyperref}
\hypersetup{colorlinks=true,linkcolor=blue,citecolor=magenta}
\usepackage{booktabs}
\DeclareMathOperator{\sign}{sign}

\begin{document}
	\renewcommand{\baselinestretch}{1.15}
	
	\title{Extended thermodynamical topology of black hole}
	
	\preprint{}
	
	\author{Shan-Ping Wu, Si-Jiang Yang, Shao-Wen Wei \footnote{Corresponding author. E-mail: weishw@lzu.edu.cn}}
	
	\affiliation{$^{1}$Key Laboratory of Quantum Theory and Applications of MoE, Lanzhou Center for Theoretical Physics,
		Key Laboratory of Theoretical Physics of Gansu Province,
		Gansu Provincial Research Center for Basic Disciplines of Quantum Physics, Lanzhou University, Lanzhou 730000, China\\	
		$^{2}$Institute of Theoretical Physics $\&$ Research Center of Gravitation,
		School of Physical Science and Technology, Lanzhou University, Lanzhou 730000, China}

	\begin{abstract}
			Thermodynamical topology has emerged as a powerful framework for classifying the thermodynamical behavior of black holes. Three distinct yet complementary topological invariants have been employed to characterize black hole phases, spinodal curves, and critical points in black hole thermodynamics. In this work, we develop a unified framework that integrates these three topological approaches and introduce the concept of extended thermodynamical topology, providing a clear physical interpretation. As a first step, we apply this framework to black holes in Einstein gravity, systematically elucidating their phase structure in terms of topological invariants. We then extend our analysis to black holes in 7-dimensional Lovelock gravity, where novel thermodynamic phenomena naturally emerge from the topological perspective. Moreover, we explore the connection between critical exponents and the extended thermodynamical topology, uncovering a correspondence between the zeros of the $k$-th order vector field and the associated critical exponents. Our study demonstrates that extended thermodynamical topology offers a robust and fine-grained framework for analyzing and classifying black hole phase transitions.
	\end{abstract}
	
	
	\maketitle
	\newpage
	\noindent\rule{180 mm}{0.5pt}
	\tableofcontents
	\noindent\rule{180 mm}{0.5pt}
	
	\newpage

	\section{Introduction}\label{Sec_Intro}
	
	Since the introduction of Bekenstein-Hawking entropy and Hawking radiation~\cite{Bekenstein:1973entropy,Hawking:1975Particle}, black hole thermodynamics has been central to our understanding of black holes, offering profound insights into the pursuit of a quantum theory of gravity. Early studies, both via quantum field theory in curved spacetime~\cite{Hawking:1975Particle,Birrell:1982QFTCurvedSpace} and through the on-shell gravitational path integral~\cite{Hartle:1976Path,Gibbons:1976ActionIntegral}, established that black holes exhibit thermodynamic behavior. It was later discovered that black holes, similar to ordinary matter, admit a rich thermodynamic phase structure and can undergo phase transitions. In particular, Schwarzschild black holes in anti-de Sitter (AdS) spacetime exhibit two distinct phases (small black holes and large black holes), along with the Hawking-Page phase transition between the thermal AdS and large black hole phases~\cite{Hawking:1982HP}. Similarly, Reissner-Nordstr\"om AdS (RN-AdS) black holes display small, intermediate, and large black hole phases, with a coexistence curve and a critical point that closely resemble the liquid-gas phase transition in van der Waals fluid~\cite{Chamblin:1999ChargeAdS,Chamblin:1999ChargeAdS2,Kubiznak:2012PV}. Moreover, the fact that black hole entropy scales with the horizon area reveals the presence of novel microscopic degrees of freedom intrinsic to gravitational systems, imposing stringent constraints on any consistent theory of quantum gravity, which must recover the Bekenstein-Hawking entropy in the semi-classical limit~\cite{Strominger:1996Microscopic,Rovelli:1996Loop}. The universality of this area law has also played a central role in the development of gauge/gravity duality~\cite{tHooft:1973,tHooft:1993,Susskind:1994,Maldacena:1997LargeN,Witten:1998AdSholo}, and the Hawking-Page transition corresponds to the confinement/deconfinement transition of quark-antiquark pairs in the dual gauge theory~\cite{Witten:1998Confine}. Recently, inspired by the holographic principle, black hole thermodynamics has been reformulated within a holographic context, including the first laws and their relation to conformal field theories, leading to renewed interest and active investigation~\cite{Kubiznak:2016,Visser:2021,Cong:2021,Ahmed:2023,Yang:2024krx}.
	
	With the development of various gravitational theories, a wide range of black hole solutions beyond the Kerr-Newman (KN) family has been discovered, and their thermodynamic properties have been extensively studied. Notable examples include black holes in Lovelock gravity~\cite{Boulware:1985String,Wheeler:1985LoveLock,Cai:2003LoveLock} and quasi-topological gravity~\cite{Myers:2010BH,Oliva:2010cubic,Hennigar:2017,Bueno:2019QTG1,Bueno:2019QTG2,Bueno:2022QTG}, among others. Given that many of these black hole solutions display qualitatively similar thermodynamic behavior, it is both natural and necessary to develop a systematic framework for classifying black hole thermodynamics. To this end, the concept of thermodynamical topology was introduced~\cite{Wei:2022Top1}. Within this framework, at a fixed ensemble temperature, each black hole solution is treated as a topological defect in the space of thermodynamic parameters. The sign of its associated winding number encodes information on thermodynamic stability: a positive winding number indicates a locally stable branch, while a negative winding number signals local instability. The sum of all winding numbers defines a thermodynamical topological number, an invariant that classifies black hole thermodynamics into three categories, with numbers of $+1$, $0$, and $-1$. For instance, Schwarzschild black holes carry a topological number of $-1$, RN black holes have $0$, and RN-AdS black holes exhibit $+1$. This classification scheme has been successfully applied to a broad class of black hole solutions, including rotating and rotating AdS black holes~\cite{Wu:2022rotating,Wu:2023rotatingAdS}.
	
	Further developments have refined thermodynamical topology into four distinct classes by incorporating the stability of the innermost small black hole branch~\cite{Wei:2024UniverTop}. When the total topological number is zero, the innermost state may be either stable or unstable, denoted as $W^{0+}$ or $W^{0-}$, respectively. For nonzero topological numbers of $1$ or $-1$, the innermost branch is necessarily stable or unstable, and is denoted by $W^{1+}$ or $W^{1-}$, respectively. This refined classification enables a more detailed analysis of both high- and low-temperature regimes and discriminates between solutions with the same zero topological number, such as Schwarzschild AdS and RN black holes, which still exhibit qualitatively different thermodynamic behaviors. Nonetheless, this scheme does not encompass all black hole solutions. In particular, some systems undergo topological number transitions, triggered by changes in the boundary behavior of the parameter space. Examples include multiply charged black holes in gauged supergravity~\cite{Duff:1999SuperGravity,Behrndt:1998SuperGravity,Wu:2024Rcharge} and four-dimensional dyonic AdS black holes in quasi-topological electrodynamics~\cite{Liu:2019QuasiTopologicalEM,Chen:2024dyonic}. To account for these cases, a novel class $W^{0- \leftrightarrow 1+}$ has been introduced, along with two additional sub-classes, $\bar{W}^{1+}$ and $\hat{W}^{1+}$~\cite{Wu:2024NovelTop}. These refinements further broaden and deepen the thermodynamical topology framework, offering a more complete classification of black hole thermodynamics.
	
	The preceding discussion outlined the thermodynamical topology used to characterize black hole phases. In black hole thermodynamics, critical points play a central role in the phase diagram, marking the termination of coexistence curves and encoding the character of phase transitions. To capture these features, a topological description of critical points has been introduced~\cite{Wei:2021Top2}. In this framework, each critical point is assigned a winding number, leading to two distinct classes: conventional and novel ones. For instance, RN-AdS black holes exhibit only conventional critical points, while Born-Infeld black holes possess both conventional and novel types. In addition to the classification of critical points, topological techniques have also been employed to study the Davies-type phase transition~\cite{Bhattacharya:2024TopDavies}, which is characterized by the divergence in the heat capacity and corresponds to the spinodal curve in the temperature-entropy ($T_H-S$) plane~\cite{Davies:1977,Davies:1989,Lousto:1993,Lousto:1994jd,Muniain:1995ih}.
	
	The three topological numbers introduced above respectively characterize black hole phases, Davies-type transition points, and critical points, capturing complementary aspects of black hole thermodynamics. Although both Schwarzschild AdS and RN-AdS black holes share the same topological number (the first topological number) of $+1$, only the RN-AdS black hole exhibits a phase transition from a small black hole to a large black hole. This distinction is reflected in their respective topological numbers associated with critical points. These results suggest that a complete thermodynamic classification should incorporate all three topological numbers and may require the inclusion of additional invariants. To address this issue, we construct a unified formalism that integrates the three established topological numbers and introduces new invariants, thereby providing a comprehensive framework for the classification of black hole thermodynamic behaviors.
	
	In this paper, we advance the framework of thermodynamical topology to provide a unified characterization of black hole thermodynamic properties. In Sec.~\ref{Sec_Review}, we review the three established topological numbers relevant to black hole thermodynamics. In Sec.~\ref{Sec_GFunction}, we reformulate these numbers within a unified formalism and introduce additional topological invariants. Sec.~\ref{Sec_TopEinstein} applies this extended framework to black holes in Einstein gravity, while Sec.~\ref{Sec_7lovelock} examines 7-dimensional spherically symmetric Lovelock black holes as a representative case. In Sec.~\ref{Sec_CriticalExponents}, we explore the connection between  extended thermodynamical topology and the associated critical exponents. Sec.~\ref{Sec_discussion} concludes with a summary of our results and final remarks.
	
	\section{Review of Thermodynamical Topology}\label{Sec_Review}
	
	Topology generally captures the global structure and intrinsic properties of a system, independent of its local details. In black hole thermodynamics, topological methods are employed to reveal global features of the system, thereby bypassing the intricacies of detailed analytical computations. This approach enables a systematic and robust classification of black hole thermodynamic behavior using topological invariants. A range of thermodynamical topologies has been proposed to describe various aspects of black hole thermodynamics~\cite{Wei:2022Top1,Wei:2021Top2,Bhattacharya:2024TopDavies,Wei:2024UniverTop,Wu:2024NovelTop,Fan:2022bsq,Yerra:2022coh,Fang:2022rsb}. In this work, we focus on several topological invariants derived from Duan's $\phi$-mapping topological current theory~\cite{Duan:1979,Duan:1984}, which identify and characterize the zeros of associated vector fields. A brief review of this formalism is presented in Sec.~\ref{Sec_ReviewDuanTopologicalCurrent}. In Sec.~\ref{Sec_ThreeTypes}, we revisit three types of thermodynamic topological numbers that characterize black hole phase states, Davies-type transition points, and critical points, respectively.
	
	\subsection{A Brief Review of Duan's $\phi$-mapping Topological Current Theory}\label{Sec_ReviewDuanTopologicalCurrent}
	
	To characterize the zeros of a two-dimensional vector field $\phi$, we introduce the associated unit vector $n^a = \phi^a/\left\| \phi \right\|$, enabling the application of Duan's $\phi$-mapping topological current theory~\cite{Duan:1979,Duan:1984}. The corresponding topological current is defined as
	\begin{equation}
		j^{\mu}=\frac{1}{2\pi}\epsilon ^{\mu \nu \rho}\epsilon _{ab}\partial _{\mu}n^a\partial _{\rho}n^b,
	\end{equation}
	where $\epsilon_{ab}$ and $\epsilon_{\mu \nu \rho}$ are the antisymmetric Levi-Civita symbols in two and three dimensions, respectively. The indices are raised and lowered by using the Euclidean metric. It can be verified that this current is conserved, i.e., $\partial_{\mu} j^{\mu} = 0$. Furthermore, utilizing the identities $\Delta _{\phi}\ln \left\| \phi \right\| =2\pi \delta ^2\left( \phi \right) $ and $\epsilon ^{ab}J^{\mu}\left( x,\phi \right) =\epsilon ^{\mu \nu \rho}\partial _{\mu}\phi ^a\partial _{\rho}\phi ^b$, we obtain
	\begin{equation}
		j^{\mu}=\delta ^2\left( \phi \right) J^{\mu}\left( x,\phi \right) .
	\end{equation}
The Dirac delta function is nonzero at the zeros of the vector field $\phi$. Denoting the isolated zeros by $z^i$, the density of the topological current becomes
	\begin{equation}
		j^0=\sum_i{w_i \delta ^2\left( x-z_i \right)}, \quad w_i=\beta _i\eta _i.
	\end{equation}
	The positive Hopf index $\beta_i$ measures the number of loops in the vector space as $x$ encircles the zero $z_i$, while the Brouwer degree $\eta_i$ is given by the sign of the Jacobian determinant evaluated at the zero, i.e., $ \eta _i=\sign(\left. J(x,\phi ) \right|_{z_i}) $. Integrating over a small neighborhood $\Sigma _{\delta}\left( z_i \right)$ surrounding the zero $z_i$ yields,
	\begin{eqnarray}
		w_i&=&\int_{\Sigma _{\delta}\left( z_i \right)}{j^0d^2x}\nonumber\\
&=&\frac{1}{2\pi}\int_{\Sigma _{\delta}\left( z_i \right)}{\epsilon ^{\mu \nu \rho}\epsilon _{ab}\partial _{\mu}n^a\partial _{\nu}n^bd^2x}\nonumber\\
&=&\frac{1}{2\pi}\Delta \Omega = \frac{1}{2\pi}\int_{\partial \Sigma _{\delta}\left( z_i \right)}{d\Omega},
		\label{eq_wcontInt}
	\end{eqnarray}
	where we used Stokes's theorem. The argument $\Omega$ of the vector reads
	\begin{equation}
		n^1=\left\| n \right\| \cos  \Omega , \quad n^2=\left\| n \right\| \sin  \Omega. \label{omega}
	\end{equation}
	Hence, $w_i$ quantifies the total rotation (winding) of the unit vector field $n$ (or equivalently, vector field $\phi$) along the closed contour $\partial \Sigma_\delta(z_i)$ and is identified as the winding number associated with the zero at $z_i$. Extending the integral over the entire parameter space $\Sigma$ accounts for all isolated zeros, yielding the total topological number
	\begin{equation}
		W=\int_{\Sigma}{j^0d^2x} = \sum_i{w_i}.
	\end{equation}
	Once again invoking Stokes' theorem, this also admits a contour integral representation over the boundary $\partial \Sigma$,
	\begin{equation}
		W = \frac{1}{2\pi}\Delta \Omega,\quad \Delta \Omega = \int_{\partial \Sigma}{d\Omega}.
		\label{eq_QcontInt}
	\end{equation}
	Therefore, the total topological number depends solely on the behavior of the vector field $n$ (or $\phi$) along the boundary of $\Sigma$, and is insensitive to the detailed structure of the field in the interior.

 In the following discussion on the winding number, we perform contour integrals along closed elliptical curves. These specific contours take the form of
    \begin{equation}
        \left( a+A\cos \left( \sigma \right) ,b+B\sin \left( \sigma \right) \right),
        \label{eq_EllipticalCurve}
    \end{equation}
    where the curve parameter $\sigma$ ranges over the interval $(0,2\pi)$. By evaluating the contour integral along these closed paths, we can intuitively extract the angular variation of the associated vector field, thereby determining the corresponding winding number.

	\subsection{Three Types of Thermodynamic Topology of Black Holes}~\label{Sec_ThreeTypes}
	
	To characterize black hole phase states, we begin by introducing the concept of generalized off-shell free energy, in which the black hole mass and temperature are treated as independent variables, extending the standard definition of free energy~\cite{York:1986BH,Whiting:1988BH}. Specifically, the generalized free energy takes the form
	\begin{equation}
		F = M - \frac{1}{\tau}S,
		\label{eq_freeenergy}
	\end{equation}
	where $M$ denotes the black hole energy and $S$ its entropy. The parameter $\tau$ is the inverse of the ensemble temperature, which is independent of the Hawking temperature $T_H$. The relation between $M$ and $S$ follows from black hole thermodynamics and is typically expressed in terms of the horizon radius $r_h$. One can find that, for fixed $\tau$, the stationary points of the generalized free energy $F$ with respect to $r_h$ (or equivalently $S$) correspond to black hole solutions whose Hawking temperature equals 1/$\tau$. At these stationary points, the generalized free energy reduces to the standard free energy. To emphasize this, we refer to it as the on-shell free energy, denoted by $F_\text{on-shell}$. Motivated by these observations, a two-dimensional vector field is constructed as~\cite{Wei:2022Top1},
	\begin{equation}
		\phi _0\left( r_h,\Theta \right) =\left( \frac{\partial F}{\partial r_h},-\cot \Theta  \csc \Theta \right),
		\label{eq_phi0}
	\end{equation}
	where $\Theta$ is an auxiliary parameter. It can be seen that the zeros of the second component of the vector field arise solely from the condition $\Theta = \pi/2$, and therefore, the first component warrants closer attention. The zeros of the first component correspond to the condition
	\begin{equation}
		\frac{\partial M}{\partial r_h}-\frac{1}{\tau}\frac{\partial S}{\partial r_h}=0,\quad \text{or} \quad T_H=\frac{1}{\tau}.
	\end{equation}	
	Thus, the zeros of this vector field correspond to black hole solutions at inverse temperature $\tau$.
	
	Next, let us consider the Davies-type phase transition~\cite{Davies:1977,Davies:1989,Lousto:1993,Lousto:1994jd,Muniain:1995ih}, which corresponds to the divergence points of the heat capacity and, equivalently, to the spinodal curve in the $T_H-S$ diagram. To describe this transition, the following vector field is introduced as~\cite{Bhattacharya:2024TopDavies}
	\begin{equation}
		\phi _1\left( S,\Theta \right) =\left( \frac{1}{\sin \Theta}\frac{\partial}{\partial S}\left( \frac{1}{T_H} \right) ,-\frac{\cot \Theta \csc \Theta}{T_H} \right) ,
		\label{eq_phi1}
	\end{equation}
	where $T_H$ denotes the Hawking temperature and $\Theta$ again serves as an auxiliary parameter. Given that the heat capacity is defined as $T_H (\partial_S T_H)^{-1}$, the zeros of $\phi_1$ identify Davies-type critical points.
	
	Finally, the critical point, which marks the endpoint of the coexistence curve, is given by the conditions
	\begin{equation}
		\frac{\partial T_H}{\partial S}=0,\ \ \frac{\partial ^2T_H}{\partial S^2}=0.
		\label{eq_THcritical}
	\end{equation}
	For simplicity, we consider the temperature in the form $T_H = T_H(S, P)$, where $P$ denotes the thermodynamic pressure~\cite{Kastor:2009Enthalpy,Dolan:2011Pressure}. Solving $\partial_S T_H(S, P) = 0$ yields an expression $P=P(S)$. The associated vector field is then defined as~\cite{Wei:2021Top2},
	\begin{equation}
		\phi _2\left( S,\Theta \right) =\left( \frac{1}{\sin \Theta}\frac{\partial}{\partial S}T_H\left( S,P\left( S \right) \right) ,-T_H\left( S,P\left( S \right) \right) \cot \Theta \csc \Theta \right) .
		\label{eq_phi2}
	\end{equation}
	The first component is proportional to $\left. \left( \partial _{S}^{2}T_H \right) \right|_{P=P\left( S \right)}$, since the substitution $P = P(S)$ implies the first condition in Eq.~\eqref{eq_THcritical}. Thus, the zeros of $\phi_2$ correspond to thermodynamic critical points.
	
	In summary, three distinct two-dimensional vector fields have been introduced to characterize the essential thermodynamic features of black holes. It can be seen that the zeros of the first component of each vector field encode key thermodynamic properties of the black hole, while the second component is constructed from an auxiliary parameter $\Theta$, whose zeros are fixed at $\Theta = \pi/2$. On the other hand, according to Duan's $\phi$-mapping topological current theory (reviewed in Sec.~\ref{Sec_ReviewDuanTopologicalCurrent}), each zero of a given vector field is assigned a winding number,
	\begin{equation}
		w_i=\beta _i\eta _i,
	\end{equation}
	where $\beta_i$ is the Hopf index and $\eta_i$ the Brouwer degree. The sum over all such contributions gives the total topological number $W = \sum_{i} w_i$, which is a topological invariant depending only on the behavior of the vector field at the boundary of its domain. Applying this theory to the vector fields $\phi_0$, $\phi_1$, and $\phi_2$, we obtain three corresponding topological numbers that characterize black hole solutions at fixed temperature, Davies-type transition points, and critical points, respectively.
	
	Another noteworthy point is that the zeros of these vector fields reflect an underlying hierarchical structure. The vector fields $\phi_0$, $\phi_1$ and $\phi_2$ correspond to the solution sets that satisfy increasingly constrained thermodynamic conditions, namely,
	\begin{align}
		\frac{1}{\tau}&=T_H ;\label{eq_pd0T}\\
		 \frac{1}{\tau}&=T_H, \quad \partial _ST_H =0;\label{eq_pd1T}\\
		  \frac{1}{\tau}&=T_H, \quad \partial _ST_H=0, \quad \partial _{S}^{2}T_H  =0,
		  \label{eq_pd2T}
	\end{align}	
	respectively. Each successive condition imposes an additional derivative constraint on the temperature $T_H$. This opens the possibility for a unified formulation of the three previously introduced vector fields, as well as the construction of new vector fields and topological invariants that further refine the classification of black hole thermodynamics. In the next section, we discuss this in more detail.

	\section{Extended Thermodynamical Topology}\label{Sec_GFunction}
	In the preceding section, we show that the three vector fields capture distinct aspects of black hole thermodynamics. Accordingly, employing multiple topological invariants together enables a more comprehensive characterization of complex thermodynamic behavior. However, the current formulations of the three known topological numbers lack full uniformity, as demonstrated by the structural differences among the vector fields defined in Eqs.~\eqref{eq_phi0}, \eqref{eq_phi1}, and \eqref{eq_phi2}. This motivates the development of a more natural and unified framework for describing these thermodynamical topology in the context of black hole thermodynamics.
	
	We begin by recalling the three thermodynamical topologies introduced above, each associated with the thermodynamic conditions in Eqs. \eqref{eq_pd0T}, \eqref{eq_pd1T}, and \eqref{eq_pd2T}. Using the generalized free energy defined in Eq.~\eqref{eq_freeenergy}, these conditions can be equivalently recast as
	\begin{align}
		\partial _SF&=0;
		\label{eq_PSF0} \\
		\partial _SF&=0,\quad \partial _{S}^{2}F=0;
		\label{eq_PSF1}\\
		\partial _SF&=0,\quad \partial _{S}^{2}F=0,\quad \partial _{S}^{3}F=0,
		\label{eq_PSF2}
	\end{align}
	which reflect increasing degeneracy of the generalized free energy. To consider the condition in Eq.~\eqref{eq_PSF0}, we define the vector field
	\begin{equation}
		\Phi _0 \left( S,\Theta \right) =\left( \partial _S F,-\cot \Theta  \csc \Theta \right),
		\label{eq_Phi0}
	\end{equation}
	whose zeros correspond to black hole phases at fixed ensemble temperature. Similar to Eq.~\eqref{eq_phi0}, an auxiliary parameter $\Theta$ is introduced. This construction of the second component of the vector field indicates that its zero is fixed at $\Theta = \pi/2$. Consequently, when analyzing the zeros of the vector field, we primarily focus on its first component. In such cases, despite a slight abuse of terminology, we denote the zero of the vector field at $(S,\Theta) = (S_c,\pi/2)$ simply as $S = S_c$ for convenience. Nevertheless, when considering winding numbers and topological numbers, the second component of the vector field and the auxiliary parameter $\Theta$ remain essential. Now, let us consider the winding number associated with each zero of the vector field $\Phi_0$. It is determined by the sign of the Jacobian determinant of the vector field mapping $\Phi_0$, which is given by
	\begin{equation}
		\left| \begin{matrix}
			\partial _S\Phi_0 ^S&		\partial _S\Phi_0 ^{\Theta}\\
			\partial _{\Theta}\Phi_0 ^S&		\partial _{\Theta}\Phi_0 ^{\Theta}\\
		\end{matrix} \right|_{\left( S,\Theta \right) =\left( S_c,\pi /2 \right)}=\left. \partial _{S}^{2}F \right|_{S=S_c}.
	\end{equation}
	Unless otherwise stated, all Jacobian determinants and winding numbers in the following expressions are evaluated at the corresponding zeros of the vector fields. For notational simplicity, we omit the explicit indication of the evaluation points. On the other hand, the heat capacity $C$ satisfies the relation
	\begin{equation}
			\partial_{S}^2 F = T_H C^{-1}.
	\end{equation}
	Thus, we can find
	\begin{equation}
		w^{(0)} = \sign(C^{-1}),
	\end{equation}
	indicating that positive winding numbers correspond to locally stable branches, while negative values signal local instability. This relation directly links the local thermodynamic stability to the winding number.
	
	For the second pair of conditions in Eq.~\eqref{eq_PSF1}, we treat the first equation, $\partial_S F = 0$, as a constraint and define the associated vector field as
	\begin{equation}
		\Phi _1\left( S,\Theta \right) =\left( \left. \partial _{S}^{2}F \right|_{\partial _SF=0},-\cot \Theta \csc \Theta \right) ,
		\label{eq_Phi1}
	\end{equation}
	whose zeros correspond to stationary points of the curve $T_H-S$. Thus, these zeros are identified as Davies-type transition points, where the heat capacity diverges. The winding number at each zero of the vector field $\Phi_1(S,\Theta)$ is determined by the sign of the Jacobian determinant $\partial_S \Phi_1^S$, which corresponds to differentiating $\partial_S^2 F$ with respect to $S$ under the constraint $\partial_S F = 0$. To evaluate this derivative in a constrained setting, we introduce a Lagrange multiplier $\lambda$ and consider the combination
	\begin{equation}
		\partial _{S}^{2}F+\lambda \partial _SF.
	\end{equation}
	Taking the derivative with respect to $S$ and evaluating at the zeros, the second term vanishes, yielding
	\begin{equation}
		\partial _{S}^{3}F=\partial _S\left( T_H C^{-1} \right) =T_H \partial _S\left( C^{-1} \right).
	\end{equation}
	Note that this Jacobian determinant is evaluated at the zeros of the vector field $\Phi_1$. Consequently, the heat capacity itself diverges at these points, while its inverse vanishes. The winding number is therefore governed by the behavior of the inverse heat capacity near its zeros. Specifically, the winding number is given by
	\begin{equation}
		w^{(1)} = \sign(\partial_S (C^{-1})),
	\end{equation}
	so a positive (negative) winding number indicates a transition from a locally unstable (stable) to a stable (unstable) phase as the entropy increases through the zero of the vector field.
	
	For the third set of conditions in Eq.~\eqref{eq_PSF2}, which includes three equations, we impose the first two constraints, $\partial_S F = 0$ and $\partial_S^2 F = 0$, and define the vector field
	\begin{equation}
		\Phi _2\left( S,\Theta \right) =\left( \left. \partial _{S}^{3}F \right|_{\left\{ \partial _SF=0,\partial _{S}^{2}F=0 \right\}},-\cot \Theta \csc \Theta \right),
		\label{eq_Phi2}
	\end{equation}
	whose zeros identify the critical points. The winding number at each zero is determined by the sign of the Jacobian determinant $\partial_S \Phi_2^S$. To evaluate this, we introduce Lagrange multipliers $\lambda_1$ and $\lambda_2$, and compute
	\begin{equation}
		\partial _S\left( \partial _{S}^{3}F+\lambda _1\partial _SF+\lambda _2\partial _{S}^{2}F \right) =T_H\partial _S^2 \left( C^{-1} \right),
	\end{equation}
	which leads to the winding number
	\begin{equation}
		w^{(2)} = \sign(\partial_S^2 (C^{-1})).
	\end{equation}
	Expanding the inverse heat capacity near a zero of $\Phi_2$ at $S = S_c$, we find
	\begin{align}
		\frac{1}{C\left( S \right)}=&\frac{1}{C\left( S_c \right)}+\left. \frac{\partial}{\partial S}\left( \frac{1}{C\left( S \right)} \right) \right|_{S=S_c}\left( S-S_c \right) +\frac{1}{2}\left. \frac{\partial ^2}{\partial S^2}\left( \frac{1}{C\left( S \right)} \right) \right|_{S=S_c}\left( S-S_c \right) ^2+\mathcal{O} \left( \left( S-S_c \right) ^3 \right)
		\\
		=&\frac{1}{2} \left. \frac{\partial ^2}{\partial S^2}\left( \frac{1}{C\left( S \right)} \right) \right|_{S=S_c}\left( S-S_c \right) ^2+\mathcal{O} \left( \left( S-S_c \right) ^3 \right),
	\end{align}
	since the first two terms vanish by construction. This shows that the leading-order behavior is quadratic, and that the sign of the second derivative determines the winding number. A positive (negative) winding number indicates that both neighboring black hole phases around $S_c$ are locally stable (unstable), respectively.

	From the hierarchical structure of the vector fields defined in Eqs.~\eqref{eq_Phi0}, \eqref{eq_Phi1}, and \eqref{eq_Phi2}, a clear nested pattern emerges. More generally, we can define the $k$-th order vector field as
	\begin{equation}
		\Phi _k\left( S,\Theta \right) =\left( \left. \partial _{S}^{k+1}F \right|_{\left\{ \partial _{S}^{m}F=0,\ m=1,\ \dots,\ k. \right\}},-\cot \Theta \csc \Theta \right),
		\label{eq_Phikdef}
	\end{equation}
	whose zeros correspond to simultaneous solutions of the conditions
	\begin{equation}
		\partial_S F = 0,\quad \partial_S^2 F = 0, \quad \dots\ , \quad \partial_S^{k+1} F = 0.
		\label{eq_PSkF}
	\end{equation}
	The winding number associated with each zero is determined by the sign of the Jacobian determinant for $\Phi_k$, which can be evaluated by introducing Lagrange multipliers $\lambda_1$, $\lambda_2$, $\dots$, $\lambda_k$, and computing
	\begin{equation}
		\partial _S\left( \partial _{S}^{k+1}F+\lambda _1\partial _SF+\lambda _2\partial _{S}^{2}F+\dots+\lambda _k\partial _{S}^{k}F \right) =T_H\partial _{S}^{k}\left( C^{-1} \right),
		\label{eq_Jocabik}
	\end{equation}
	where we have used the fact that both $C^{-1}$ and all lower-order derivatives of $F$ vanish at the zero of $\Phi_k$. It follows that the winding number of $\Phi_k$ is given by
	\begin{equation}
		w^{(k)} = \sign(\partial_S^k (C^{-1})).
		\label{eq_wkpdkC}
	\end{equation}
	This construction proceeds recursively by successively incorporating higher-order derivatives. Accordingly, we refer to $\Phi_k$ as the $k$-th order vector field, $w^{(k)}$ as the $k$-th order winding number, and $W^{(k)}$ as the corresponding $k$-th order topological number. The collection $\{ W^{(k)}, k = 0, 1, 2, \dots \}$ defines the extended thermodynamic topology. To analyze the local behavior near the zero of $\Phi_k$ at $S = S_c$, we expand the inverse heat capacity as,
	\begin{align}
		\frac{1}{C\left( S \right)}=&\frac{1}{C\left( S_c \right)}+\left. \frac{\partial}{\partial S}\left( \frac{1}{C\left( S \right)} \right) \right|_{S=S_c}\left( S-S_c \right) +\dots+\frac{1}{k!} \left. \frac{\partial ^k}{\partial S^k}\left( \frac{1}{C\left( S \right)} \right) \right|_{S=S_c}\left( S-S_c \right) ^k+\mathcal{O} \left( \left( S-S_c \right) ^{k+1} \right) \nonumber
		\\
		=&\frac{1}{k!}\left. \frac{\partial ^k}{\partial S^k}\left( \frac{1}{C\left( S \right)} \right) \right|_{S=S_c}\left( S-S_c \right) ^k+\mathcal{O} \left(\left( S-S_c \right) ^{k+1} \right) ,
		\label{eq_Cleading}
	\end{align}
	where all lower-order terms vanish as a consequence of the conditions in Eq.~\eqref{eq_PSkF}. As shown in Eq.~\eqref{eq_wkpdkC}, the winding number is determined by the sign of the leading coefficient in this expansion, and hence directly reflects the local thermodynamic stability. Considering that $S = S_c$ is a simple root, the Jacobian determinant is nonzero and this zero carries a nonvanishing winding number. For odd $k$, a winding number of $+1$ ($-1$) indicates that the system transitions from a locally unstable (stable) to a stable (unstable) branch as $S$ increases through $S_c$, reflecting a change in stability. For even $k > 0$, a winding number of $+1$ ($-1$) implies that both sides of $S_c$ correspond to locally stable (unstable) phases, and thus no stability change occurs. In the special case $k = 0$, the winding number directly reflects the stability of the phase at $S_c$: $+1$ for a locally stable phase, and $-1$ for an unstable one.
	
	In the foregoing discussion, including that of Eqs. \eqref{eq_Phi0}, \eqref{eq_Phi1}, and \eqref{eq_Phi2}, we have implicitly assumed that each zero of the vector field $\Phi_k$ is a simple root. However, if the root is degenerate, additional subtleties arise.  Suppose that $S = S_c$ is a zero of $\Phi_k$ with multiplicity $m + 1$. Then, by construction, $S=S_c$ is also a zero of the higher-order vector fields $\Phi_{k+1}$, $\Phi_{k+2}$, $\dots$,  $\Phi_{k+m}$. Expanding the inverse heat capacity around this point, the leading non-vanishing term takes the form
	\begin{equation}
		\frac{1}{C\left( S \right)}=\frac{1}{(k+m)!}\left. \frac{\partial ^{k+m}}{\partial S^{k+m}}\left( \frac{1}{C\left( S \right)} \right) \right|_{S=S_c}\left( S-S_c \right) ^{k+m}+\mathcal{O} \left( \left( S-S_c \right) ^{k+m+1} \right),
	\end{equation}
	indicating that $\Phi_{k+m}$ possesses a simple zero at $S_c$. In such cases, the thermodynamic behavior can be interpreted within the simple root framework introduced earlier. This observation highlights the importance of introducing higher-order vector fields and their associated topological numbers in the thermodynamic analysis of black holes. Such a framework provides a more refined physical characterization, particularly in situations where lower-order descriptions fail to capture essential features. Moreover, a multiple root may be regarded as a superposition of coincident simple roots, each contributing a winding number of $\pm1$. The $k$-th order winding number associated with such a multiple zero is given by the sum of these individual contributions and can be computed via a contour integral of $\Phi_k$ around the degenerate point, as expressed in Eq.~\eqref{eq_wcontInt}. As an illustrative example, consider the RN-AdS black hole at its critical point, where the AdS radius and temperature take their critical values. In this case, $\Phi_0$ admits a triple zero, while $\Phi_2$ possesses a simple zero. The total winding number of $\Phi_0$ is $+1$, corresponding to three coincident simple zeros contributing $+1$, $-1$, and $+1$, respectively. Another important point is that the physical interpretation of the topological number remains valid, as it characterizes the sum of the winding numbers of all zeros and reflects a global property of black hole thermodynamics.
	
	On the other hand, when interpreting thermodynamic behavior through the topological framework, the role of the Lagrange multipliers $\lambda_1$, $\lambda_2$, $\cdots$, $\lambda_k$ warrants further clarification. In constructing the vector field $\Phi_k(S,\Theta)$, the conditions $ \partial _{S}F=0$, $ \partial _{S}^2 F=0$, $\cdots$, $\partial_S^k F = 0$ are treated as independent constraints. These constraints typically eliminate $k$ independent variations from $\partial_S^{k+1} F$. By introducing $k$ Lagrange multipliers, one formally treats these parameters as independent during differentiation. As compensation, the derivatives of the total expression
	\begin{equation}
		 \partial _{S}^{k+1}F+\lambda _1\partial _SF+\lambda _2\partial _{S}^{2}F+\dots+\lambda _k\partial _{S}^{k}F,
	\end{equation}
	with respect to these $k$ parameters must vanish, thereby determining the Lagrange multipliers. In the previous analysis, the explicit values of the Lagrange multipliers do not affect the final value of the Jacobian determinant, as shown in Eq.~\eqref{eq_Jocabik}, since the conditions in Eq.~\eqref{eq_PSkF} are satisfied at the zero $S = S_c$ of $\Phi_k$. Nevertheless, it is important to ensure that none of the Lagrange multipliers diverges at $S = S_c$, as any such divergence would invalidate the physical interpretation of the winding number. In practice, however, the Lagrange multipliers $\lambda_1$, $\lambda_2$, $\cdots$, $\lambda_k$ often remain finite at the zero of $\Phi_k$, and we therefore do not pursue this issue further unless an explicit divergence arises in certain special cases.

	\section{Thermodynamical Topology of Black Holes in Einstein Gravity}\label{Sec_TopEinstein}
	
	Einstein gravity constitutes one of the simplest and most fundamental frameworks for describing gravitation. In the asymptotically flat electrovacuum setting, the most general stationary black hole solution is uniquely described by the KN spacetime. This black hole is characterized by its mass $M$, angular momentum $J$ (often parametrized by the specific angular momentum $a \equiv J/M$), and electric charge $Q$. The associated thermodynamic quantities are given by
	\begin{equation}
		M=\frac{1}{2r_h}\left( r_{h}^{2}+a^2+Q^2 \right) ,\quad S=\pi \left( r_{h}^{2}+a^2 \right) ,\quad J=\frac{1}{2r_h}a\left( r_{h}^{2}+a^2+Q^2 \right) ,
	\end{equation}
	where $r_h$ denotes the radius of the event horizon and $a$ is the specific angular momentum. In the context of black hole thermodynamics, the mass $M$ is interpreted as a thermodynamic potential and treated as a function of entropy $S$, angular momentum $J$, and electric charge $Q$, i.e., $M = M(S, J, Q)$. The first law of black hole thermodynamics then takes the form
	\begin{equation}
		\delta M=T_H\delta S+\Omega _H\delta J+\Phi \delta Q,
	\end{equation}
	where $T_H$, $\Omega_H$, and $\Phi$ are the Hawking temperature, angular velocity, and electric potential of the horizon, respectively,
	\begin{equation}
		T_H=\frac{r_{h}^{2}-a^2-Q^2}{4\pi r_h\left( r_{h}^{2}+a^2 \right)},\quad \Omega _H=\frac{a}{r_{h}^{2}+a^2},\quad \Phi =\frac{Qr_h}{r_{h}^{2}+a^2}.
	\end{equation}
	The generalized free energy is given by
	\begin{equation}
		F=M-\frac{1}{\tau}S,
		\label{eq_FreeEnergyEinstein}
	\end{equation}
	where $\tau$ is the inverse ensemble temperature. The KN black hole reduces to the Schwarzschild solution when $Q = 0$ and $J = 0$, to the RN black hole when $J = 0$ and $Q \neq 0$, and to the Kerr black hole when $J \neq 0$ and $Q = 0$. Building on the topological framework developed in Sec.~\ref{Sec_GFunction}, we analyze the thermodynamic topology for both the limiting cases and the full KN family to uncover their broader thermodynamic implications.
	
	\subsection{Case I: Schwarzschild black hole}
	In the case of the Schwarzschild black hole, representing the simplest scenario, the generalized free energy takes the form
	\begin{equation}
		F=\frac{\sqrt{S}}{2\sqrt{\pi}}-\frac{S}{\tau}.
		\label{eq_FreeEnergySBH}
	\end{equation}
	Consequently, the zeroth order vector field on the parameter space $(S,\Theta)$ is given by
	\begin{equation}
		\Phi _0=\left( \frac{1}{4\sqrt{\pi S}}-\frac{1}{\tau},-\cot \Theta  \csc \Theta \right),
	\end{equation}
	with domain $\left\{ ( S,\Theta) \middle| 0<S<\infty ,0<\Theta <\pi \right\} $. Examining the asymptotic behavior of each component yields,
	\begin{align}
		&\left. \Phi _{0}^{S} \right|_{S\rightarrow 0}=\frac{1}{4\sqrt{\pi S}}+\mathcal{O} \left( 1 \right) , \quad \left. \Phi _{0}^{S} \right|_{S\rightarrow \infty}=-\frac{1}{\tau}+\mathcal{O} \left( S^{-1/2} \right),
		\\
		&\left. \Phi _{0}^{\Theta} \right|_{\Theta \rightarrow 0}=-\frac{1}{\Theta ^2}+\mathcal{O} \left( 1 \right) ,\quad \left. \Phi _{0}^{\Theta} \right|_{\Theta \rightarrow \pi}=\frac{1}{(\Theta -\pi )^2}+\mathcal{O} \left( 1 \right).
	\end{align}
	In the $(S,\Theta)$ parameter space, the vector field exhibits the following behavior along four segments of the boundary: along $S=0$, it points in the positive $S$-direction; along $\Theta = 0$, in the negative $\Theta$-direction; along $S = \infty$, in the negative $S$-direction; and along $\Theta = \pi$, in the positive $\Theta$-direction. Traversing these segments in sequence defines a single counterclockwise contour, during which the field itself undergoes a net $2\pi$ clockwise rotation. According to Eq.~\eqref{eq_QcontInt}, this yields a topological number $W^{(0)} = -1$, confirming that the Schwarzschild black hole resides in an unstable thermodynamic phase. Since the second component of the vector field is identical in every construction, we henceforth focus on the first component when extracting the thermodynamical topology. Imposing the constraint $\partial_S F = 0$, we obtain the first order vector field from $\partial_S^2 F$,
	\begin{equation}
		\Phi _1=\left( -\frac{1}{8\sqrt{\pi}S^{3/2}},-\cot \Theta  \csc \Theta \right).
	\end{equation}
	Although the inverse ensemble temperature $\tau$ explicitly cancels when the generalized free energy~\eqref{eq_FreeEnergySBH} is differentiated twice with respect to $S$, the constraint $\partial_S F = 0$ remains essential, as it encodes the relationship between $\tau$ and the entropy $S$. From the expression of $\Phi_1$, we observe that it admits no zeros, and the associated topological number vanishes. This confirms the absence of a thermodynamic phase transition for the Schwarzschild black hole.
	
\begin{figure}[h]
		\begin{center}
			\subfigure[ \ $\Phi_0 (S,\Theta)$
			\label{FIG_SBHphi0}]{\includegraphics[width=5cm]{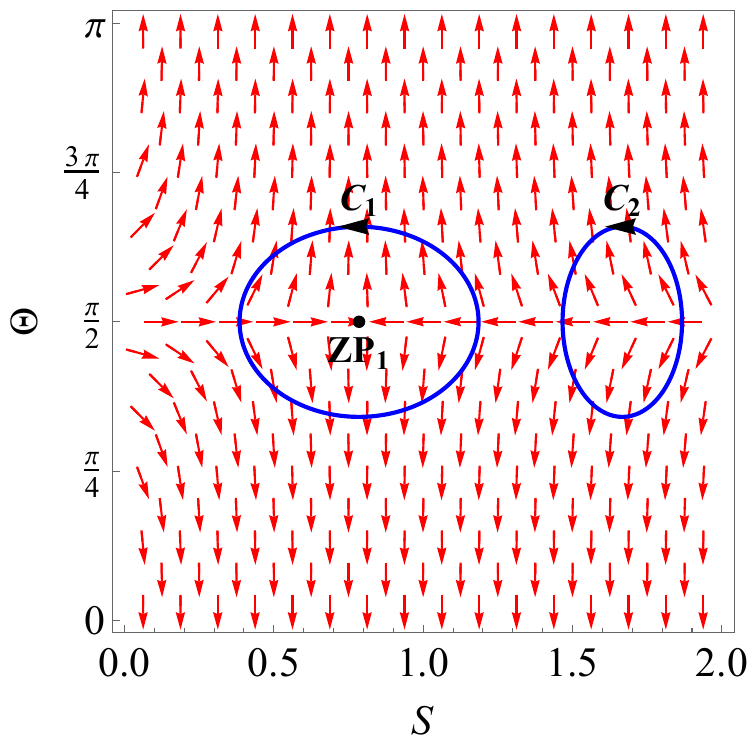}}
			\subfigure[\ $\Phi_1 (S,\Theta)$
			\label{FIG_SBHphi1}]{\includegraphics[width=5cm]{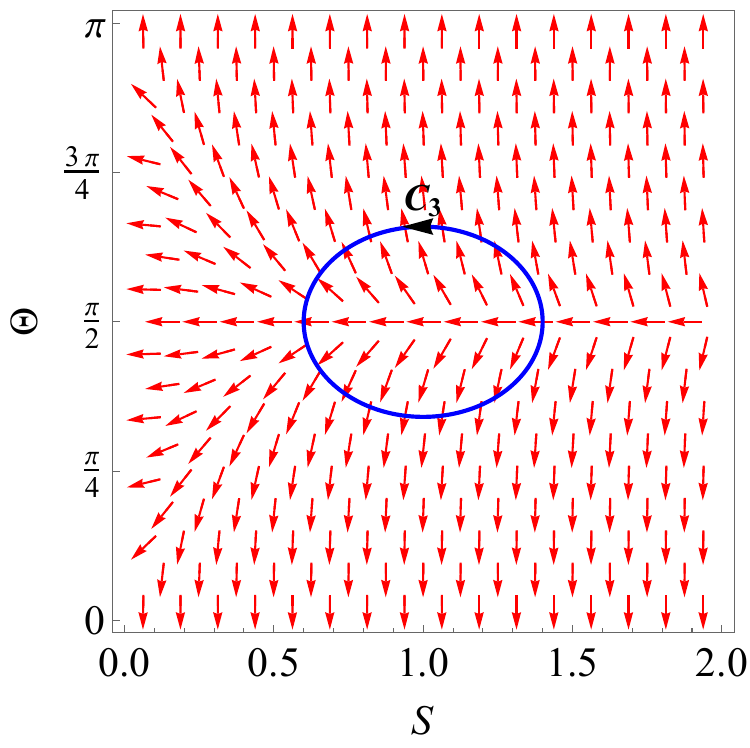}}
		\end{center}
		\caption{Behavior of the normalized vector fields of $\Phi_0$ and $\Phi_1$ for Schwarzschild black hole. (a) The normalized vector field of $\Phi_0$ with $\tau = 2\pi$, and the zero point $ZP_1$ is located at $(\pi/4,\pi/2)$. (b) The normalized vector field of $\Phi_1$. In both sub-figures, the closed curves labeled $C_1$, $C_2$, and $C_3$, each taking the form given in Eq.~\eqref{eq_EllipticalCurve}, are also depicted.}
		\label{FIG_SBHphi}
	\end{figure}

\begin{figure}
		\begin{center}
			{\includegraphics[width=5cm]{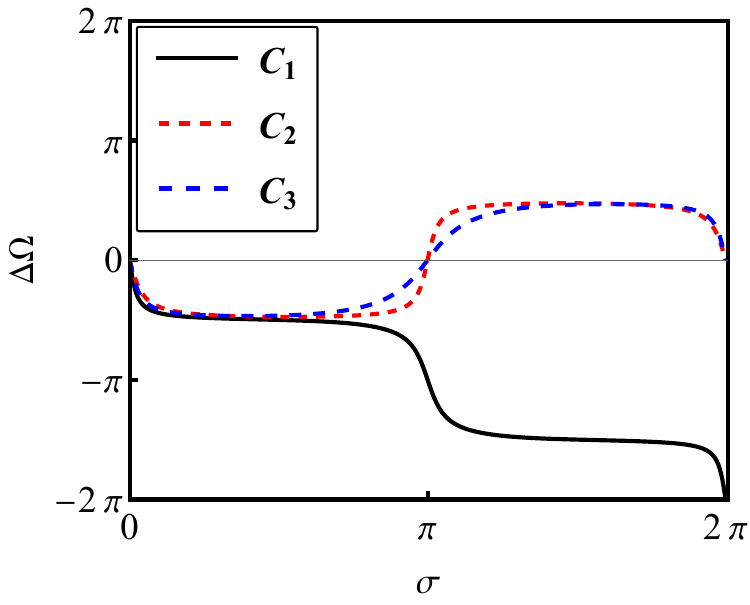}}
		\end{center}
		\caption{Variation of the angle of the vector fields for Schwarzschild black hole. $C_1$ and $C_2$ are the closed curves in Fig.~\ref{FIG_SBHphi0}, while $C_3$ corresponds to the closed curve in Fig.~\ref{FIG_SBHphi1}. The vertical axis, $\Delta \Omega$, denotes the accumulated angular change of the vector field along the respective curves, and the horizontal axis, $\sigma$, is the curve parameter, ranging over the interval $(0,2\pi)$.}
		\label{FIG_SBHContour}
	\end{figure}

	To provide a more intuitive understanding, illustrative vector field plots are presented. Fig.~\ref{FIG_SBHphi} shows the vector field diagrams of $\Phi_0$ and $\Phi_1$, along with a closed contour. The point $ZP_1$ denotes a zero of the vector field $\Phi_0$, while $C_1$, $C_2$ and $C_3$ represent three distinct closed contours, each following the form specified in Eq.~\eqref{eq_EllipticalCurve}. Fig.~\ref{FIG_SBHContour} shows the variation of the argument of the vector along these contours. Here, $\Delta \Omega$ denotes the change in the angle of the vector field along the corresponding closed contour, and $\sigma$ is the curve parameter ranging from $0$ to $2\pi$. Specifically, $\Delta \Omega$ is given by Eq.~\eqref{eq_wcontInt}, and it is related to the winding number $w$ through $w=\Delta \Omega /(2\pi)$. Intuitively, one can see that the winding number $w^{(0)}$ associated with the zero $ZP_1$ of the vector field $\Phi_0$ is $-1$, contributing $-1$ to the corresponding topological number $W^{(0)}$.
	
	\subsection{Case II: Reissner-Nordstr\"om black hole}
\begin{figure}[h]
		\begin{center}
			\subfigure[\ $\Phi_0 (S,\Theta)$
			\label{FIG_RNBHphi0}]{\includegraphics[width=5cm]{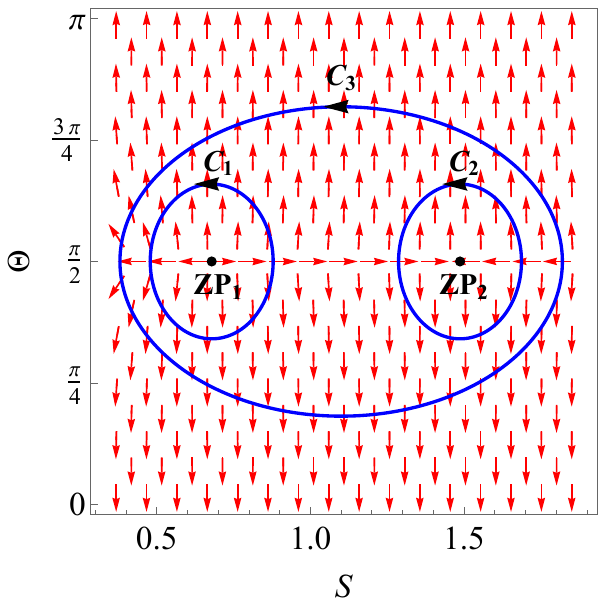}}
			\quad  \quad
			\subfigure[\ $\Phi_1 (S,\Theta)$
			\label{FIG_RNBHphi1}]{\includegraphics[width=5cm]{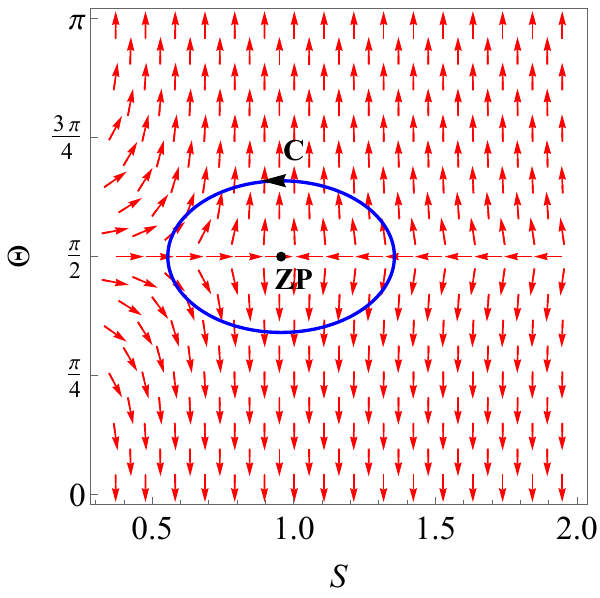}}
		\end{center}
		\caption{Behavior of the normalized vector fields of $\Phi_0$ and $\Phi_1$ for RN black hole. (a) The normalized vector field of $\Phi_0$ with $\tau = 11$ and $Q=1/\pi$. The zero points $ZP_1$, $ZP_2$ are approximately located at $(0.679,\pi/2)$ and $(1.487,\pi/2)$, respectively. (b) The normalized vector field of $\Phi_1$ with $Q=1/\pi$. The zero point $ZP$ is located at $(3/\pi,\pi/2)$. In both sub-figures, the closed curves labeled $C_1$, $C_2$, $C_3$, and $C$, each taking the form given in Eq.~\eqref{eq_EllipticalCurve}, are also depicted.}
		\label{}
	\end{figure}

\begin{figure}[h]
		\begin{center}
		 {\includegraphics[width=5 cm]{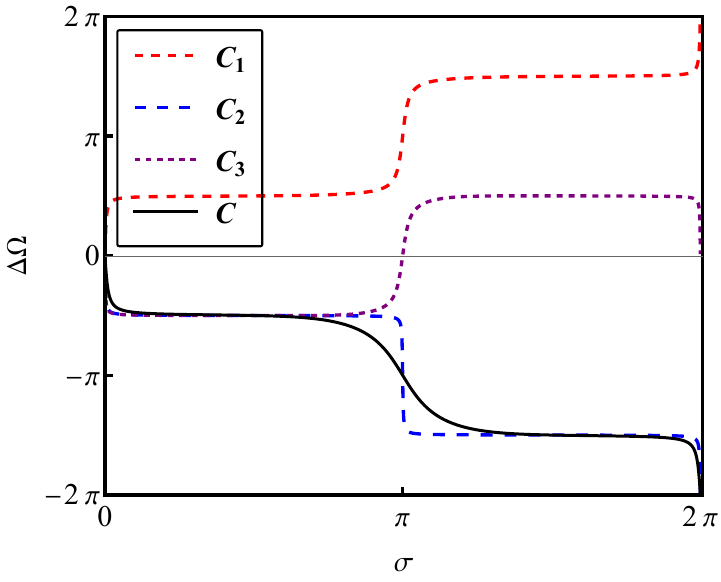}}
		\end{center}
		\caption{Variation of the angle of the vector fields for RN black hole. $C_1$, $C_2$ and $C_3$ are the closed curves in Fig.~\ref{FIG_RNBHphi0}, while $C$ corresponds to the closed curve in Fig.~\ref{FIG_RNBHphi1}. The vertical axis, $\Delta \Omega$, denotes the accumulated angular change of the vector field along the respective curves, and the horizontal axis, $\sigma$, is the curve parameter, ranging over the interval $(0,2\pi)$.
		}
		\label{FIG_RNBHC}
	\end{figure}

	When the electric charge is included, the generalized free energy becomes
	\begin{equation}
		F=\frac{S+\pi Q^2}{2\sqrt{\pi S}}-\frac{S}{\tau}.
	\end{equation}
	Note that the RN black hole horizon radius attains a minimum in the extremal limit, so the entropy is bounded from below by its extremal value, i.e., $S \in[S_{m0},\infty)$, with $S_{m0} = \pi  Q^2 $. Examining the asymptotic behavior of the first component of the vector field $\Phi_0$, we find
	\begin{equation}
		\left. \Phi _{0}^{S} \right|_{S\rightarrow S_{m0}}=-\frac{1}{\tau}+\mathcal{O} \left( \left( S-S_{m0} \right) \right),\quad \left. \Phi _{0}^{S} \right|_{S\rightarrow \infty}=-\frac{1}{\tau}+\mathcal{O} \left( S^{-1/2} \right),
	\end{equation}
	indicating that the zeroth order topological number vanishes. This implies that locally stable and unstable black hole phases must occur in pairs. The behavior of $\Phi_0$ is illustrated in Fig.~\ref{FIG_RNBHphi0}. The variation of the argument of vector $\Omega$ along the closed contour shown in Fig.~\ref{FIG_RNBHphi0} is plotted in Fig.~\ref{FIG_RNBHC}. As evident from these figures, $\Phi_0$ admits two zeros with winding numbers $+1$ and $-1$, respectively, confirming that the total topological number $W^{(0)}$ vanishes.
	
\begin{figure}[h]
		\begin{center}
			\subfigure[ \ $T_H-S$ curve
			\label{FIG_RNBHTH}]{\includegraphics[width=5.25cm]{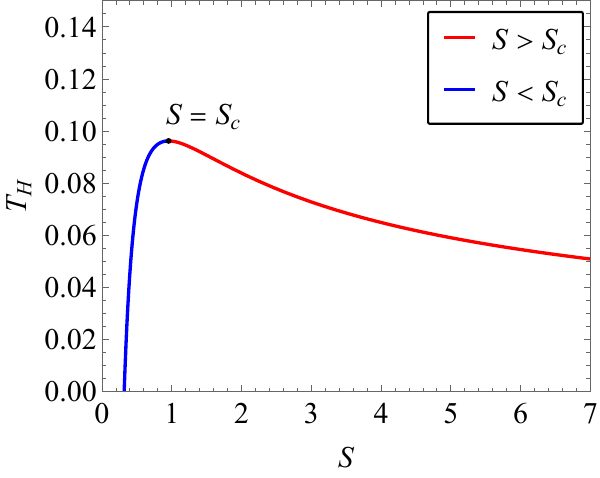}}
			\quad  \quad
			\subfigure[ \ $F_\text{on-shell}-T_H$ curve
			\label{FIG_RNBHFE}]{\includegraphics[width=5cm]{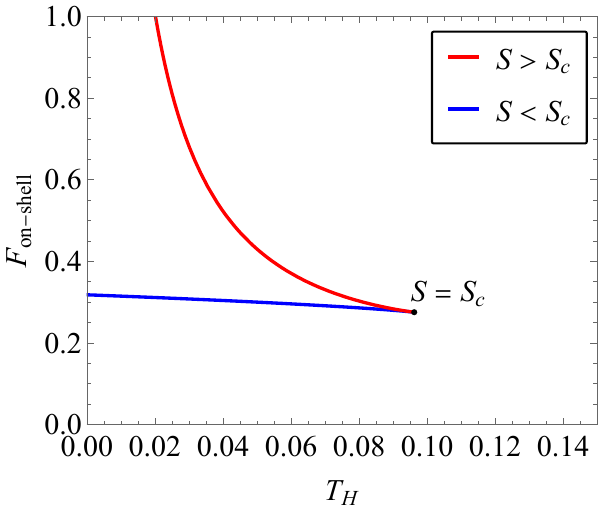}}
		\end{center}
		\caption{Thermodynamic characteristic curves of the RN black hole. (a) The relation between entropy and temperature. (b) Behavior of the on-shell free energy with respect to temperature. The charge parameter is set to $Q = \pi^{-1}$. In both sub-figures, the red curve ($S>S_c$) represents the unstable large black hole phase, and the blue curve ($S<S_c$) corresponds to the locally stable small black hole phase. The two branches are separated at the entropy $S = S_c$.
		}
		\label{FIG_RNBHSTTF}
	\end{figure}

	Imposing the condition $\partial_S F =0$ determines the inverse temperature $\tau$ as a function of $S$ and $Q$. Since $\tau>0$, the entropy $S$ has a lower bound $S_{m1}$, which coincides with $ S_{m0} = \pi  Q^2$. Taking an additional derivative, $\partial_S^2 F$, yields the first order vector component. Its asymptotic behavior is given by
	\begin{equation}
		\left. \Phi _{1}^{S} \right|_{S\rightarrow S_{m1}}=\frac{1}{4\pi ^2\left| Q \right|^3}+\mathcal{O} \left( \left( S-S_{m0} \right) \right),\quad \left. \Phi _{1}^{S} \right|_{S\rightarrow \infty}=-\frac{1}{8\sqrt{\pi}S^{3/2}}+\mathcal{O} \left( S^{-5/2} \right).
	\end{equation}
	This behavior indicates that the first order topological number $W^{(1)}$ is equal to $-1$. The vector field $\Phi_1$ is shown in Fig.~\ref{FIG_RNBHphi1}, and the variation of its angular component $\Omega$ along the associated closed contour is presented in Fig.~\ref{FIG_RNBHC}. The presence of a simple zero of $\Phi_1$ signals a transition in the black hole phase and its thermodynamic character at $S = S_c$: for $S > S_c$, the black hole is locally thermodynamically unstable, whereas for $S < S_c$, it is locally stable. On the other hand, explicit calculations show that $\Phi _1\sim \left( 3\pi Q^2-S \right) $, implying that $S_c = 3\pi Q^2$. The heat capacity takes the form
	\begin{equation}
		C_Q=T\left( \frac{\partial S}{\partial T} \right) _Q=\frac{2S\left( S-\pi Q^2 \right)}{3\pi Q^2-S}
	\end{equation}
	which is consistent with the behavior inferred from the topological number. To further illustrate this behavior, we also present the $T_H-S$ and $F_{\text{on-shell}}-T_H$ diagrams in Fig.~\ref{FIG_RNBHSTTF}. These plots clearly show that the critical entropy $S_c$ separates the black hole phase into two distinct branches: one locally thermodynamically stable and the other unstable.
	
	Finally, enforcing the conditions $\partial _SF=0$ and $\partial _{S}^{2}F=0$ yields the relations
	\begin{equation}
	 	\tau =6\sqrt{\pi S}, \quad Q^2=\frac{S}{3\pi}.
	\end{equation}
	Substituting these into $\partial_S^3 F$ leads to the second order vector field
	\begin{equation}
		\Phi _2=\left( -\frac{1}{8\pi S^{5/2}},-\cot \Theta  \csc \Theta \right) .
	\end{equation}
	This vector field has no zeros, confirming the absence of any critical points in the thermodynamic of the RN black hole. This conclusion aligns with the thermodynamic behavior depicted in Fig.~\ref{FIG_RNBHTH} and Fig.~\ref{FIG_RNBHFE}.
	
	\subsection{Case III: Kerr black hole}

	Including the angular momentum, the generalized free energy becomes
	\begin{equation}
		F=\sqrt{\frac{\pi J^2}{S}+\frac{S}{4\pi}}-\frac{S}{\tau},
	\end{equation}
	and, as in the charged case, the entropy is bounded from below by the extremal limit $S_{m0}=2\pi \left| J \right|$. The asymptotic behavior of the zeroth order vector field $\Phi_0$ is given by
	\begin{equation}
		\left. \Phi _{0}^{S} \right|_{S\rightarrow S_{m0}}=-\frac{1}{\tau}+\mathcal{O} \left( \left( S-S_{m0} \right) \right) , \quad
		\left. \Phi _{0}^{S} \right|_{S\rightarrow \infty}=-\frac{1}{\tau}+\mathcal{O} \left( S^{-1/2} \right),
	\end{equation}
	indicating a vanishing zeroth order topological number. This again suggests that locally stable and unstable black hole phases occur in pairs. Using the condition $\partial_S F = 0$, $\partial_S^2 F$ yields the first order vector component with asymptotic behavior
	\begin{equation}
		\left. \Phi _{1}^{S} \right|_{S\rightarrow S_{m1}}=\frac{1}{8\pi ^2\left| J \right|^{3/2}}+\mathcal{O} \left( \left( S-S_{m0} \right) \right) ,
		\quad
		 \left. \Phi _{1}^{S} \right|_{S\rightarrow \infty}=-\frac{1}{8\sqrt{\pi}S^{3/2}}+\mathcal{O} \left( S^{-5/2} \right),
	\end{equation}
	implying a first order topological number $W^{(1)}=-1$. Imposing the conditions $\partial _SF=0$ and $\partial _SF=0$ simultaneously determines the parameters as
	\begin{equation}
		\tau =2\sqrt{\left( 3+2\sqrt{3} \right) \pi S}, \quad J^2=\frac{\left( 2\sqrt{3}-3 \right)}{12\pi ^2}S^2.
	\end{equation}
	Substituting these into $\partial_S^3 F$ leads to the second order vector field
	\begin{equation}
		\Phi _2=\left( -\frac{3^{1/4}\left( 2\sqrt{3}-3 \right)}{2\sqrt{2\pi}S^{5/2}},-\cot \Theta  \csc \Theta \right),
	\end{equation}
	whose first component remains nonzero throughout the domain of $\Phi_2$. This confirms the absence of any critical point in the thermodynamics of the Kerr black hole, similar to the behavior observed for the RN solution.

	\subsection{Case IV: Kerr-Newman black hole}
	When both charge and angular momentum are included, the generalized free energy takes the form
	\begin{equation}
		F=\sqrt{\frac{4\pi ^2J^2+\left( S+\pi Q^2 \right) ^2}{4\pi S}}-\frac{1}{\tau}S,
	\end{equation}
	with the entropy bounded below by the extremal limit $S_{m0}=\pi \sqrt{4J^2+Q^4}$. The first derivative, $\partial_S F$, leads the zeroth order vector field $\Phi_0$, whose asymptotics are
	\begin{equation}
		\left. \Phi_0^S \right|_{S\rightarrow S_{m0}}=-\frac{1}{\tau}+\mathcal{O} \left( \left( S-S_{m0} \right) \right) ,\quad \left. \Phi_0^S \right|_{S\rightarrow \infty}=-\frac{1}{\tau}+\mathcal{O} \left( S ^{-1/2} \right).
	\end{equation}
	This implies that the zeroth order topological number vanishes. As a result, for $\tau>0$, locally stable and unstable black hole phases must occur in pairs. Imposing the condition $\partial_{S} F = 0$, the first order vector field $\Phi_1$, derived from $\partial_S^2 F$, has the asymptotic form
	\begin{align}
		\left. \Phi_1^S\right|_{S\rightarrow S_{m0}}&=\frac{1}{2\sqrt{2}\pi ^2\left( 4J^2+Q^4 \right) ^{1/4}\sqrt{4J^2+Q^4+Q^2\sqrt{4J^2+Q^4}}}+\mathcal{O} \left( \left( S-S_{m0} \right) \right) ,\\
		 \ \left. \Phi_1^S \right|_{S\rightarrow \infty}&=-\frac{1}{8\sqrt{\pi}}S^{-3/2}+\mathcal{O} \left( S^{-5/2} \right).
	\end{align}
	This indicates a first order topological number $W^{(1)} = -1$. Enforcing the two conditions $\partial_S F = 0$ and $\partial_S^2 F = 0$, one obtains
	\begin{align}
		\tau &=\frac{\sqrt{6\pi}\left( 2\pi Q^2+S+\sqrt{3S^2+6\pi Q^2S+4\pi ^2Q^4} \right)}{\sqrt{\pi Q^2+\sqrt{3S^2+6\pi Q^2S+4\pi ^2Q^4}}},
		\\
		J^2 &=\frac{S}{6\pi ^2}\sqrt{3S^2+6\pi Q^2S+4\pi ^2Q^4}-\frac{3S^2+4\pi Q^2S+3\pi ^2Q^4}{12\pi ^2}.
	\end{align}
	Requiring $\tau > 0$ and $J^2 > 0$ introduces a new lower bound $S_{m2}=3\pi Q^2$. Substituting these expressions for $\tau$ and $J$ into $\partial_S^3 F$ yields the second order vector field $\Phi_2$, whose asymptotic behavior is given by
	\begin{equation}
		\left. \Phi _{2}^{S} \right|_{S\rightarrow S_{m2}}=-\frac{\left| Q \right|}{72\sqrt{3}\pi ^3Q^6}+\mathcal{O} \left( \left( S-S_{m2} \right) \right) ,\quad \left. \Phi _{2}^{S} \right|_{S\rightarrow \infty}=-\frac{3^{3/4}\left( 2-\sqrt{3} \right)}{2\sqrt{2\pi}}\frac{1}{S^{5/2}}+\mathcal{O} \left( S^{-7/2} \right).
	\end{equation}
	This indicates that the second order topological number $W^{(2)}$ vanishes, suggesting that any associated critical points must appear in pairs. Furthermore, solving the simultaneous conditions $\partial_S F = 0$, $\partial_S^2 F = 0$, and $\partial_S^3 F = 0$ yields only an unphysical solution characterized by imaginary angular momentum and vanishing entropy. This leads to two important conclusions. First, the topological classification naturally terminates at the second order vector field $\Phi_2$. Second, $\Phi_2$ has no zeros, and no critical points exist for the KN black hole.
	
	\subsection{Brief Summary of Cases}
	The hierarchy of extended thermodynamical topology is designed to capture increasingly subtle features of black hole thermodynamics, unifying the winding numbers and their physical interpretations within a coherent framework. Table~\ref{Table_EinsteinTopology} summarizes the extended topological numbers computed for each black hole family. In the asymptotically flat case, with the exception of the Schwarzschild black hole, the RN, Kerr and KN black holes share the same topological number structure. The agreement observed in the zeroth and first order topology numbers arises from the extremal limit and the asymptotic flatness of the spacetime. These two features constrain the leading asymptotic behavior of the associated vector fields, ensuring identical topological numbers at lower-orders. At higher-orders, such as for $\Phi_2$ or more generally $\Phi_k$ ($k\ge 3$), the existence and contribution of these vector fields depend on the specific form of the free energy.

	By contrast, asymptotically AdS black holes exhibit a distinctly different thermodynamical topology, as also summarized in Table~\ref{Table_EinsteinTopology}. In these spacetimes, certain locally stable phases can be globally dominant, a feature generally absent in asymptotically flat cases, where even stable phases tend to be globally subdominant. This distinction is subtly reflected in the thermodynamical topology, particularly via the zeroth order topological number that encodes the stability properties of black hole phases. The zeroth order topological number of asymptotically AdS black holes is usually $+1$, whereas for asymptotically flat black holes it tends to be $0$ or $-1$. In addition, despite possessing the same number of thermodynamic parameters as the KN black hole,  the RN-AdS and Kerr-AdS black holes require the introduction of the third order vector field, $\Phi_3$, to fully characterize their topological structure. This observation underscores that the emergence of higher-order topological numbers is governed by the underlying structure of the thermodynamic potential, rather than solely by the number of independent parameters.
	
	It is worth noting that the highest-order topological number always vanishes, reflecting the invariance of thermodynamic relations under rescaling of black hole parameters. It is customary to fix one parameter to a nonzero constant, so the corresponding zero entry in Table~\ref{Table_EinsteinTopology} can be omitted. In Sec.~\ref{Sec_7lovelock}, we adopt this scale-fixing approach to simplify the analysis of extended thermodynamical topology.
	\\
	\begin{table}[h]
		\centering
		\resizebox{6cm}{!}{
				\begin{tabular}{|c|c|c|c|c|}
					\hline
					\centering Black Hole & $\Phi_0$ & $\Phi_1$ & $\Phi_2$ & $\Phi_3$
					\\
					\hline
					\centering Schwarzschild Black Hole & $-1$ & 0 & \textemdash &  \textemdash
					\\
					\hline
					\centering RN Black Hole & 0 & $-1$ & 0 &  \textemdash
					\\
					\hline
					\centering Kerr Black Hole & 0 & $-1$ & 0 &  \textemdash
					\\
					\hline
					\centering Kerr-Newman Black Hole & 0 & $-1$ & 0 &  \textemdash
					\\
					\hline
					\centering RN-AdS Black Hole & 1 & $0$ & 1 & 0
					\\
					\hline
					\centering Kerr AdS Black Hole & 1 & $0$ & 1 & 0
					\\
					\hline
			\end{tabular}}
		\caption{Topological numbers of the extended thermodynamical topology for black holes in Einstein gravity. Each column in the table corresponds to the $k$-th order topological number for the $k$-th order vector field $\Phi_k$. The symbol ``\textemdash'' indicates that the corresponding topological number does not exist.}
		\label{Table_EinsteinTopology}
	\end{table}

	\section{Thermodynamical Topology of Black Holes in Lovelock Gravity}\label{Sec_7lovelock}
	
	Lovelock gravity is the most general second-order theory of gravitation and provides a natural extension of the Einstein-Hilbert action. In a $d$-dimensional spacetime, the Lagrangian density takes the form
	\begin{equation}
		\mathcal{L} =\frac{1}{16\pi}\sum_{k=0}^K{\hat{\alpha}_{\left( k \right)}\mathcal{L} ^{\left( k \right)}},
	\end{equation}
	where the upper limit $K$ depends on the spacetime dimension $d$ and is given by the integer part of $(d-1)/2$, i.e., $K=\lfloor (d-1)/2 \rfloor$. The coefficients $\hat{\alpha}_{(k)}$ are the Lovelock coupling constants, and each term $\mathcal{L}^{(k)}$ corresponds to the $2k$-dimensional Euler density, defined by
	\begin{equation}
		\mathcal{L} ^{\left( k \right)}=\frac{1}{2^k}\delta _{c_1d_1 \dots c_kd_k}^{a_1b_1 \dots a_kb_k}{R_{a_1b_1}}^{c_1d_1} \dots {R_{a_kb_k}}^{c_kd_k},
	\end{equation}
	where $\delta$ denotes the generalized antisymmetric Kronecker delta function and the expression involves $k$ Riemann curvature tensors. To obtain a static, spherically symmetric black hole solution, one adopts the metric ansatz
	\begin{equation}
		ds^2=-f\left( r \right) dt^2+\frac{dr^2}{f\left( r \right)}+r^2d\Omega _{d-2}^{2},
	\end{equation}
	where $d\Omega_{d-2}^2$ denotes the line element of the unit $(d-2)$-sphere with volume $\Sigma_{d-2}$. The gravitational field equations can be reduced to a single polynomial equation for the metric function $f(r)$~\cite{Boulware:1985String,Wheeler:1985LoveLock,Cai:2003LoveLock},
	\begin{equation}
		\sum_{k=0}^K{\alpha _k\left( \frac{1-f\left( r \right)}{r^2} \right) ^k=\frac{16\pi M}{\left( d-2 \right) \Sigma _{d-2}r^{d-1}}},
		\label{eq_Lovelockfr}
	\end{equation}
	where $M$ is the Arnowitt-Deser-Misner mass of the black hole. The rescaled Lovelock coupling constants $\alpha_k$ are related to the original couplings $\hat{\alpha}_{(k)}$ by
	\begin{equation}
		\alpha _0=\frac{\hat{\alpha}_{\left( 0 \right)}}{\left( d-1 \right) \left( d-2 \right)},\quad \alpha_1 =  \hat{\alpha}_{\left( 1 \right)}, \quad
		\alpha _k=\hat{\alpha}_{\left( k \right)}\prod_{n=3}^{2k}{\left( d-n \right)}.
	\end{equation}
	From the polynomial equation \eqref{eq_Lovelockfr}, the thermodynamic properties of the black hole can be analyzed from the horizon geometry~\cite{Cai:2003LoveLock}. For a black hole with event horizon radius $r_h$, one obtains
	\begin{align}
		M=&\frac{\left( d-2 \right) \Sigma _{d-2}}{16\pi}\sum_{k=0}^K{\alpha _kr_{h}^{d-1-2k}},
		\\
		T_H =&\frac{f^{\prime}\left( r_h \right)}{4\pi}=\frac{1}{4\pi r_h}\left( \sum_{k=1}^K{k\alpha _kr_{h}^{-2k}} \right) ^{-1}\sum_{k=0}^K{\left( d-2k-1 \right) \alpha _kr_{h}^{-2k}},
		\\
		S=&\frac{\left( d-2 \right) \Sigma _{d-2}}{4}\sum_{k=1}^K{\frac{k\alpha _kr_{h}^{d-2k}}{d-2k}}.
	\end{align}
	The first law of black hole thermodynamics holds in the form $\delta M = T_H \delta S$.
	
	In the following, we focus on the case of the 7-dimensional black holes with $K = 3$, and restrict our attention to positive Lovelock coupling constants. In this setting, the thermodynamic quantities can be expressed explicitly as,
	\begin{align}
		&M=\frac{5\pi ^2}{16}\left( \alpha _0r_{h}^{6}+r_{h}^{4}+\alpha _2r_{h}^{2}+\alpha _3 \right) ,\nonumber \\
        &S=\frac{1}{12}\pi ^3\left( 3r_{h}^{5}+10\alpha _2r_{h}^{3}+45\alpha _3r_h \right),\nonumber  \\
		&T_H=\frac{r_h\left( 3\alpha _0 r_{h}^{4}+2r_{h}^{2}+\alpha _2 \right)}{2\pi \left( r_{h}^{4}+2\alpha _2r_{h}^{2}+3\alpha _3 \right)}.
		\label{eq_THlovelock7}
	\end{align}
	These satisfy the first law $\delta M = T_H \delta S$. Since all Lovelock coupling constants are taken to be positive, the Hawking temperature is strictly positive for all $r_h > 0$, meaning that $r_h$ ranges over $(0, \infty)$. Moreover, as the entropy $S$ is a continuous and monotonic function of positive $r_h$, the asymptotic behavior of $k$-th order vector field $\Phi_k$ on the boundary of the $(S, \Theta)$ parameter space can equivalently be parameterized by the variables $(r_h, \Theta)$.
	
	In Lovelock gravity, the generalized free energy is given by
	\begin{equation}
		F=M-\frac{1}{\tau}S=\frac{5\pi ^2}{16}\left( \alpha _0r_{h}^{6}+r_{h}^{4}+\alpha _2r_{h}^{2}+\alpha _3 \right) -\frac{1}{\tau}\frac{1}{12}\pi ^3\left( 3r_{h}^{5}+10\alpha _2r_{h}^{3}+45\alpha _3r_h \right),
	\end{equation}
	Unlike in the pure Einstein case, the generalized free energy here is expressed explicitly in terms of the horizon radius $r_h$. Nevertheless, due to the one-to-one correspondence between $S$ and $r_h$, the generalized free energy can also be regarded as an implicit function of the entropy. On the other hand, the thermodynamic quantities in Lovelock gravity exhibit a scaling invariance, as evident in the expressions in Eq.~\eqref{eq_THlovelock7}. To simplify the analysis, we employ this invariance and set $\alpha_3 = 1$ without loss of generality.
	
	The $S$-component of the zeroth order vector field, defined as $\partial_S F$, exhibits the following asymptotic behavior,
	\begin{equation}
		\left. \Phi _{0}^{S} \right|_{r_h\rightarrow 0}=-\frac{1}{\tau}+\mathcal{O} \left( r_h \right) ,\quad \left. \Phi _{0}^{S} \right|_{S\rightarrow \infty}=\frac{3\alpha _0}{2\pi}r_h+\mathcal{O} \left( 1 \right).
	\end{equation}
	This behavior indicates that the zeroth order topological number is $W^{(0)} = 1$. Consequently, the system admits at least one thermodynamically stable black hole phase, and the number of locally stable phases exceeds that of unstable phases by one. Imposing the on-shell condition $\Phi_0^S =0 $ yields the inverse temperature
	\begin{equation}
		\tau =\frac{2\pi \left( r_{h}^{4}+2\alpha _2r_{h}^{2}+3\alpha _3 \right)}{r_h\left( 3\alpha _0r_{h}^{4}+2r_{h}^{2}+\alpha _2 \right)},
	\end{equation}
	which exactly reproduces the inverse of Hawking temperature as given in Eq.~\eqref{eq_THlovelock7}.
	
	Next, we consider the first order vector field $\Phi_1$. Its $S$-component exhibits the asymptotic behavior
	\begin{equation}
		\left. \Phi _{1}^{S} \right|_{r_h\rightarrow 0}=\frac{2\alpha _2}{45\pi ^4}+\mathcal{O} \left( r_h \right) ,\quad \left. \Phi _{1}^{S} \right|_{r_h\rightarrow \infty}=\frac{6\alpha _0}{5\pi ^4r_{h}^{4}}+\mathcal{O} \left( r_h^{-5} \right).
	\end{equation}
	This behavior indicates that the first order topological number vanishes, i.e., $W^{(1)} = 0$, which implies that Davies-type critical points appear in pairs.
	
	Imposing the combined conditions $\partial_S F = 0$ and $\partial_S^2 F = 0$ yields
	\begin{equation}
			\tau =\frac{\pi \left( r_{h}^{4}+6\alpha _2r_{h}^{2}+15 \right)}{2r_h\left( r_{h}^{2}+\alpha _2 \right)},\quad \alpha _0=\frac{2r_{h}^{6}-\alpha _2r_{h}^{4}+2\left( \alpha _{2}^{2}-9 \right) r_{h}^{2}-3\alpha _2}{3r_{h}^{4}\left( r_{h}^{4}+6\alpha _2r_{h}^{2}+15 \right)}.
			\label{eq_lovelockcon2}
	\end{equation}
	The requirement that $\tau > 0$ and $\alpha_0 > 0$ imposes a lower bound on $r_h$, denoted by $r_{h2}$, which is determined as the unique positive root of
	\begin{equation}
		2{r_{h2}}^6-\alpha _2{r_{h2}}^4+2\left( \alpha _{2}^{2}-9 \right) {r_{h2}}^2-3\alpha _2=0.
		\label{eq_rh2eq}
	\end{equation}
	For the second order vector field $\Phi_2$, the $S$-component exhibits the asymptotic behavior,
	\begin{align}
			\left. \Phi _{2}^{S} \right|_{r_h\rightarrow r_{h2}}&=\frac{32\left( {r_{h2}}^6-3\alpha _2{r_{h2}}^4+\left( 6\alpha _{2}^{2}-45 \right) {r_{h2}}^2-15\alpha _2 \right)}{25\pi ^7r_{h2}\left( {r_{h2}}^4+2\alpha _2{r_{h2}}^2+3 \right) ^3\left( {r_{h2}}^4+6\alpha _2{r_{h2}}^2+15 \right)}+\mathcal{O} \left( r_h-r_{h2} \right) ,
			\label{eq_phi2Lovelock}
			\\
			 \left. \Phi _{2}^{S} \right|_{r_h\rightarrow \infty}&=\frac{32}{25\pi ^7r_{h}^{11}}+\mathcal{O} \left( r_h^{-12} \right).
	\end{align}
	Using Eq.~\eqref{eq_rh2eq}, which defines $r_{h2}$, one can show that the leading term in Eq.~\eqref{eq_phi2Lovelock} is negative. This implies that the second order topological number is $W^{(2)} = 1$.
	
\begin{figure}[h]
		\begin{center}
			\subfigure[\label{FIG_LovelockTF1}]{\includegraphics[width=5cm]{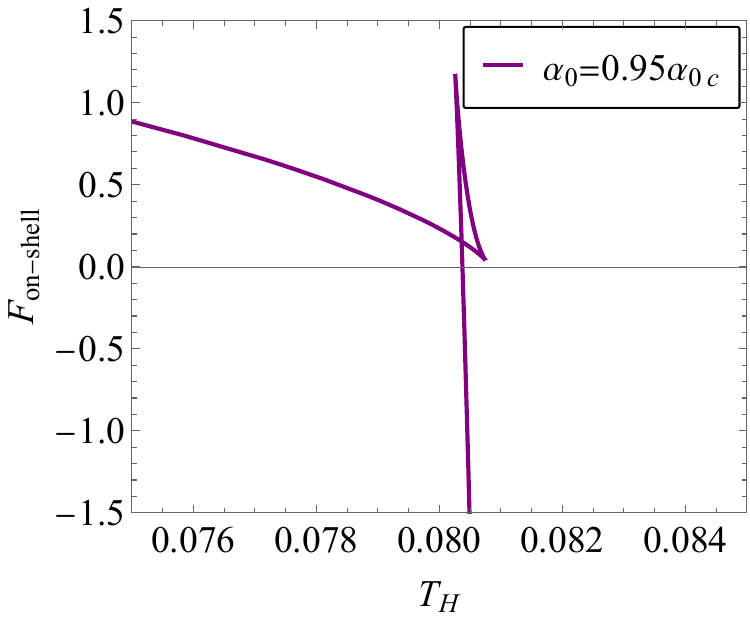}}
			\quad
			\subfigure[\label{FIG_LovelockTF2}]{\includegraphics[width=5cm]{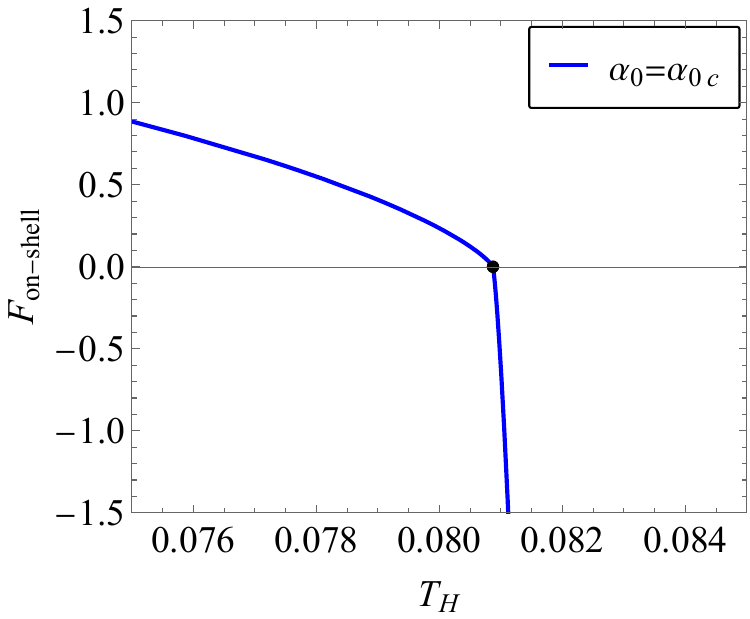}}
			 \quad
			\subfigure[\label{FIG_LovelockTF3}]{\includegraphics[width=5cm]{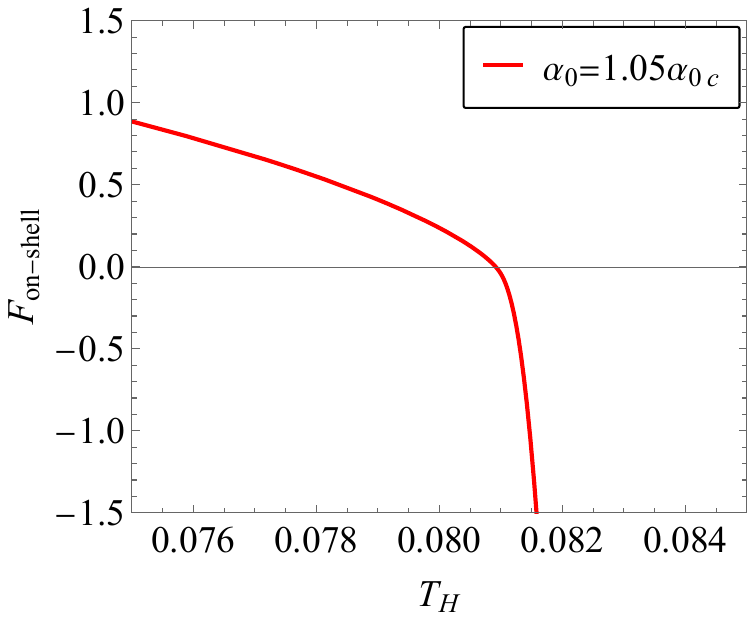}}
		\end{center}
		\caption{On-shell free energy $F_\text{on-shell}$ versus Hawking temperature $T_H$ with $\alpha _{2}=\alpha _{2c}$ and $\alpha_3 = 1$. (a) $\alpha _{0} = 0.95 \alpha _{0c}$. (b) $\alpha _{0} = \alpha _{0c}$. (c) $\alpha _{0} = 1.05 \alpha _{0c}$. A swallowtail structure is present for $\alpha_0 < \alpha_{0c}$, indicating a first order phase transition. As $\alpha_0 \ge \alpha_{0c}$, the swallowtail disappears, signaling the approach to or beyond criticality. The critical point is marked by a black dot in Fig.~\ref{FIG_LovelockTF2}, located at $T = T_{Hc}\equiv1/( 2\pi 15^{1/4} )$ and $F_\text{on-shell}=0$.}
		\label{FIG_THFLovelock}
	\end{figure}

	Imposing the conditions $\partial_S F = 0$, $\partial_S^2 F = 0$, and $\partial_S^3 F = 0$, one obtains two distinct branches of solutions for the parameters $\tau$, $\alpha_0$, and $\alpha_2$. To fully characterize the topological structure of the system, these branches must be analyzed separately.	The first branch is given by
	\begin{align}
		\tau =&\frac{\pi \left( -15+9r_{h}^{4}-\sqrt{15}\sqrt{-r_{h}^{8}+78r_{h}^{4}+15} \right)}{8r_{h}^{3}}, \nonumber \\
		\alpha _0=&\frac{5r_{h}^{8}+54r_{h}^{4}-15+\sqrt{15}\left( r_{h}^{4}+1 \right) \sqrt{-r_{h}^{8}+78r_{h}^{4}+15}}{36r_{h}^{6}\left( r_{h}^{4}-15 \right)}, \nonumber \\
		\alpha _2=&\frac{3r_{h}^{4}+15-\sqrt{15}\sqrt{-r_{h}^{8}+78r_{h}^{4}+15}}{12r_{h}^{2}}.
		\label{eq_Lovelockbranch1}
	\end{align}
	The requirement that all three parameters remain positive constrains the horizon radius to the interval
	\begin{equation}
		r_h\in \left( \sqrt{3}\sqrt[4]{5}, \sqrt[4]{16\sqrt{6}+39}~ \right].
		\label{eq_Phi31rhrange}
	\end{equation}
	Correspondingly, for the first branch of the third order vector field $\Phi_{3,1}$, the asymptotic behavior at the boundaries of this interval is
	\begin{align}
		\left. \Phi _{3,1}^{S} \right|_{r_h\rightarrow \sqrt{3}\sqrt[4]{5}}&=\frac{1}{1728000\pi ^{10}}+\mathcal{O} \left( \left( r_h-\sqrt{3}\sqrt[4]{5} \right) \right),\\
		 \left. \Phi _{3,1}^{S} \right|_{r_h\rightarrow \sqrt[4]{16\sqrt{6}+39}}&=\frac{683\sqrt{6}+1673}{20\left( 3\sqrt{6}+8 \right) ^7\left( 42\sqrt{6}+103 \right) \pi ^{10}}+\mathcal{O} \left( \left( r_h-\sqrt[4]{16\sqrt{6}+39} \right) ^{1/2} \right).
	\end{align}
	This indicates that the vector field $\Phi_{3,1}$ associated with the first branch contributes no topological number, yielding $W^{(3,1)} = 0$. For the second branch, the solutions are
	\begin{align}
		\tau =&\frac{\pi \left( -15+9r_{h}^{4}+\sqrt{15}\sqrt{-r_{h}^{8}+78r_{h}^{4}+15} \right)}{8r_{h}^{3}}, \nonumber \\
		\alpha _0=&\frac{5r_{h}^{8}+54r_{h}^{4}-15-\sqrt{15}\left( r_{h}^{4}+1 \right) \sqrt{-r_{h}^{8}+78r_{h}^{4}+15}}{36r_{h}^{6}\left( r_{h}^{4}-15 \right)}, \nonumber \\
		\alpha _2=&\frac{3r_{h}^{4}+15+\sqrt{15}\sqrt{-r_{h}^{8}+78r_{h}^{4}+15}}{12r_{h}^{2}}.
		\label{eq_Lovelockbranch2}
	\end{align}
	Positivity constraints restrict the horizon radius to
	\begin{equation}
			r_h\in \left( 0, \sqrt[4]{16\sqrt{6}+39}~ \right].
			\label{eq_Phi32rhrange}
	\end{equation}
	Accordingly, the second branch of the third order vector field $\Phi_{3,2}$ exhibits the following asymptotic behavior at the boundaries,
	\begin{align}
			\left. \Phi _{3,2}^{S} \right|_{r_h\rightarrow 0} &=\frac{1}{1600\pi ^{10}r_{h}^{4}}+\mathcal{O} \left(1\right) ,\\
			 \left. \Phi _{3,2}^{S} \right|_{r_h\rightarrow \sqrt[4]{16\sqrt{6}+39}} &=\frac{683\sqrt{6}+1673}{20\left( 3\sqrt{6}+8 \right) ^7\left( 42\sqrt{6}+103 \right) \pi ^{10}}+\mathcal{O} \left( \left( r_h-\sqrt[4]{16\sqrt{6}+39} \right) ^{1/2} \right).
	\end{align}
	This asymptotic behavior indicates that the second branch also yields a vanishing third order topological number, $W^{(3,2)} = 0$. However, a more refined analysis reveals the presence of a double root at $r_h = r_{hc} \equiv \sqrt[4]{15}$, associated with a pair of winding numbers, $+1$ and $-1$. In the vicinity of this critical radius $r_{hc}$, one finds
	\begin{equation}
		\left. \Phi _{3,2}^{S} \right|_{r_h\rightarrow \sqrt[4]{15}}=\frac{\left( r_h-r_{hc} \right) ^2}{1728000\sqrt{15}\pi ^{10}}+\mathcal{O} \left( \left( r_h-r_{hc} \right) ^3 \right).
	\end{equation}
	As discussed in Sec.~\ref{Sec_GFunction}, both black hole phases with $r_h > r_{hc}$ and $r_h<  r_{hc}$ are thermodynamically stable. The double root of the third order vector field may, in fact, be interpreted as a simple root of the fourth order vector field $\Phi_4$, reflecting the same underlying physical structure. Evaluating Eq.~\eqref{eq_Lovelockbranch2} at the critical point $r_h = r_{hc}$ yields the following characteristic relations among the parameters,
	\begin{equation}
		 \tau _c=2\pi 15^{1/4}\alpha _{3}^{1/4} ,\quad \alpha _{0c}=15^{-3/2}\alpha _{3}^{-1/2},\quad \alpha _{2c}=15^{1/2}\alpha _{3}^{1/2},
		 \label{eq_LoveLockalphac}
	\end{equation}
	where we have reinstated the explicit scaling dependence on the dimensional parameter $\alpha_3$. The relationship between the Hawking temperature and the on-shell free energy is depicted in Fig.~\ref{FIG_THFLovelock} for fixed parameters $\alpha_2 = \alpha_{2c}$ and $\alpha_3 = 1$. As shown by the solid blue curve in Fig.~\ref{FIG_LovelockTF2}, the black dot at $r_h = r_{hc}$ marks the transition between two distinct thermodynamically stable black hole phases. Notably, the free energy vanishes at this critical point. Furthermore, when $\alpha_0<\alpha_{0c}$ (e.g., $0.95 \alpha_{0c}$ as shown in Fig.~\ref{FIG_LovelockTF1}), the characteristic swallowtail structure appears, indicating a first order phase transition. In contrast, when $\alpha_0 > \alpha_{0c}$ (e.g., $1.05 \alpha_{0c}$ as shown in Fig.~\ref{FIG_LovelockTF3}), the system enters a supercritical regime.
	
\begin{figure}[h]
		\begin{center}
			\subfigure[ \ $\Phi_2 \left(r_h , \Theta \right)$
			\label{FIG_LoveLock7phi2}]{\includegraphics[width=5cm]{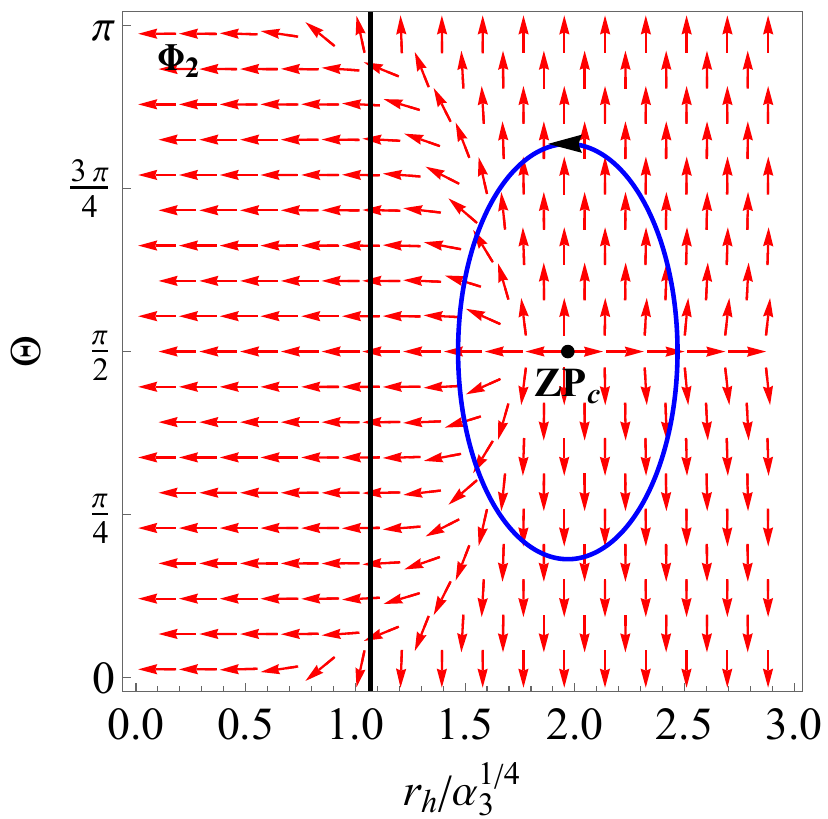}}
			\quad  \quad
			\subfigure[  \ $\Phi_{3,2} \left(r_h , \Theta \right)$
			\label{FIG_LoveLock7phi32}]{\includegraphics[width=5cm]{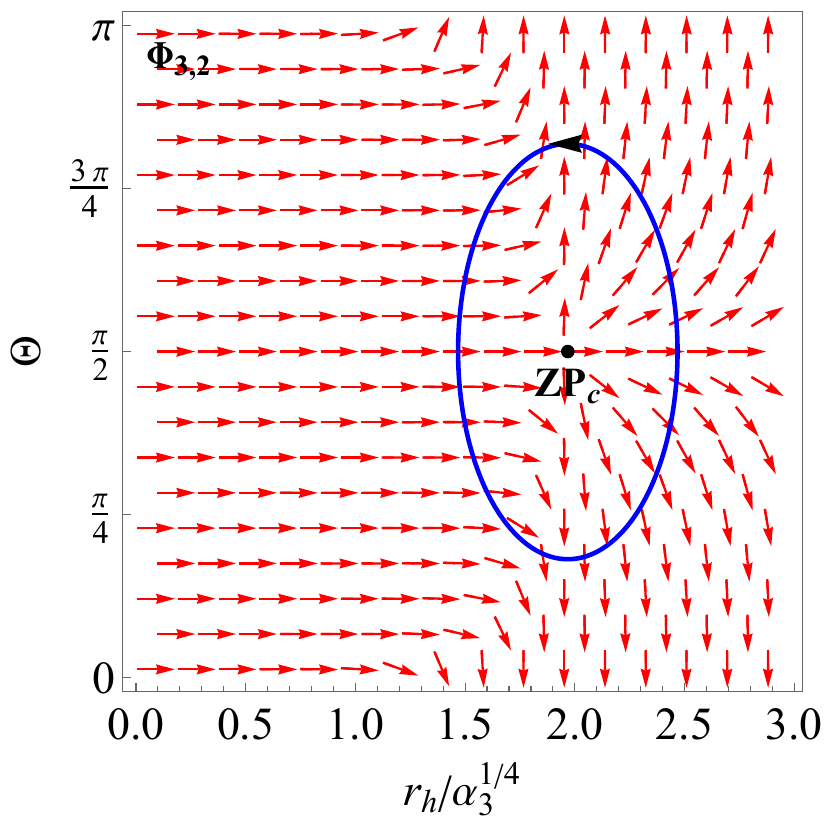}}
		\end{center}
		\caption{
			Behavior of the normalized vector fields of $\Phi_2$ and $\Phi_3$ for 7-dimensional Lovelock black hole. (a) The normalized vector field of $\Phi_2$ with $\alpha_2 = \alpha_{2c} $. (b) normalized vector field of $\Phi_{3,2}$. The vertical black line in (a) marks the boundary of the domain of $\Phi_2$, located at $r_h=r_{h2}$ (as given in Eq.~\eqref{eq_rh2eq}), i.e., $r_h/\alpha^{1/4} = 1.068$; $\Phi_2$ becomes invalid to the left of this line. To improve visibility, $\alpha _3$ is set to $1/(60\pi ^3)$ rather than 1. In both sub-figures, the two fields share a common zero point $ZP_c$ at $(15^{1/4},\pi/2)$, and two closed curves, each taking the form given in Eq.~\eqref{eq_EllipticalCurve}, are also shown.}
		\label{}
	\end{figure}

\begin{figure}
		\begin{center}
			{\includegraphics[width=5cm]{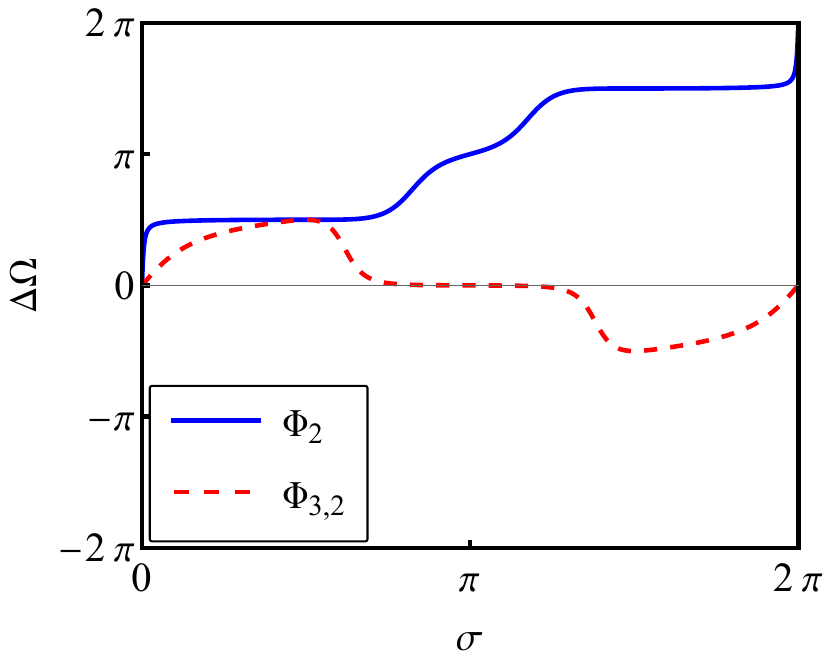}}
		\end{center}
		\caption{Variation of the angle of the vector fields for 7-dimensional Lovelock gravity. The blue solid curve shows angular variation along the closed curve in Fig.~\ref{FIG_LoveLock7phi2}, while the red dashed curve shows the angular variation along the closed curve in Fig.~\ref{FIG_LoveLock7phi32}. The vertical axis, $\Delta \Omega$, denotes the accumulated angular change of the vector field along the respective curves, and the horizontal axis, $\sigma$, is the curve parameter, ranging over the interval $(0,2\pi)$.}
		\label{FIG_LoveLock7Contour}
	\end{figure}

	To further support our analysis, we present the vector fields $\Phi_2$ and $\Phi_{3,2}$ in Fig.~\ref{FIG_LoveLock7phi2} and Fig.~\ref{FIG_LoveLock7phi32}, respectively. The corresponding angular variations along the associated closed contour are shown in Fig.~\ref{FIG_LoveLock7Contour}. The parameters are chosen according to Eq.~\eqref{eq_LoveLockalphac}, with $\alpha_3=1/(60\pi ^3)$ selected to improve the visual clarity of the plots. As illustrated, although $ZP_c$ is a common zero point of both vector fields, it contributes a second order topological number of $+1$ via $\Phi_2$.  In contrast, it makes zero contribution to the third order topological number from $\Phi_{3,2}$, consistent with our earlier analysis. The second order winding number $W^{(2)}$ of $+1$ indicates the presence of a critical point like structure, as shown in Fig.~\ref{FIG_LovelockTF2}. However, unlike a standard critical point, this winding number of $+1$ actually corresponds to a triple root, with individual contributions of $+1$, $-1$, and $+1$ from the three roots. Moreover, we find that the critical exponents near this point differ from those of a standard critical point. This behavior is discussed in detail in the next section.

\begin{figure}
		\begin{center}
			\subfigure[\label{FIG_LoveLock7alphaS}]{\includegraphics[width=4.6cm]{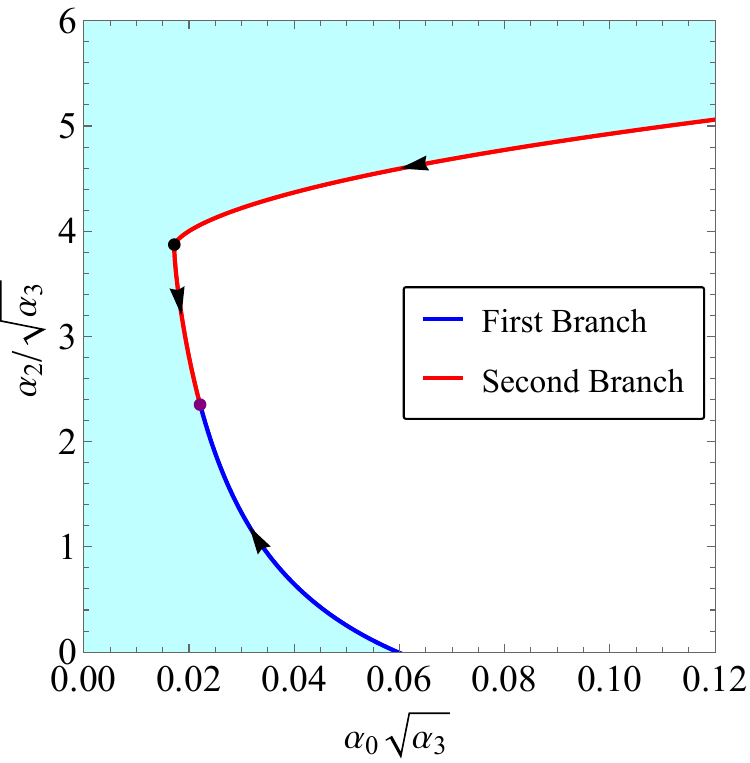}}
			\quad  \quad
			\subfigure[\label{FIG_LoveLock7alphaL}]{\includegraphics[width=5cm]{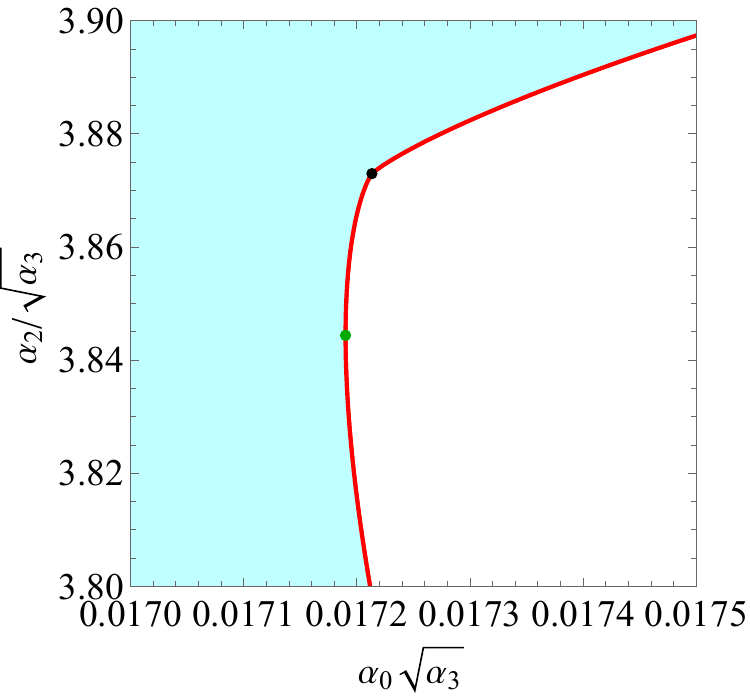}}
		\end{center}
		\caption{(a) Schematic diagram of the parameter relations for $\alpha_0$, $\alpha_2$, and $\alpha_3$ characterizing black hole phase transitions. (b) Magnified view of the region near the black dot in (a). The blue and red solid curves represent the first and second solution branches from Eqs.~\eqref{eq_Lovelockbranch1} and \eqref{eq_Lovelockbranch2}, respectively, each tracing the locus of phase-transition critical points. Arrows along the curves indicate the direction of increasing $r_h$. The black dot marks the parameters given by Eq.~\eqref{eq_LoveLockalphac}, and the purple dot corresponds to $r_h =\sqrt[4]{16\sqrt{6}+39}$, where the two branches coincide. The green dot in (b) marks the point on the curve where $\alpha_0 \sqrt{\alpha_3}$ reaches its minimum, corresponding to $r_h = \sqrt[4]{15 + 24\sqrt{2/5}}$. The lightly shaded cyan region denotes the range of parameters in which a swallowtail appears on the $F_\text{on-shell} - T_H$ diagram.}
		\label{FIG_LoveLock7alpha}
	\end{figure}

	It is worth emphasizing that the construction of the third order vector field  requires imposing the conditions $\partial_S F = 0$, $\partial_S^2 F =0 $, and $\partial_S^3 F = 0$, which yield two distinct solution branches, as given in Eqs. \eqref{eq_Lovelockbranch1} and \eqref{eq_Lovelockbranch2}. These two branches, parametrized by the horizon radius $r_h$, are shown in Fig.~\ref{FIG_LoveLock7alpha}: the blue and red solid curves represent the respective branches (i.e., loci of critical points), while the cyan shaded region marks the parameter domain where a swallowtail structure appears in the $ F_\text{on-shell} - T_H$ diagram. The black dot, located at $r_h = r_{hc}$, marks the zero of $\Phi_3$ on the second branch and corresponds to a singular point where the first derivative is continuous, while the second derivative diverges. In contrast, the purple dot denotes the point where the two branches merge smoothly, without introducing any singularity. The emergence of two distinct branches originates from choosing $r_h$ as the parametrization variable. The continuity properties at both points can also be deduced directly from Eqs.~\eqref{eq_Lovelockbranch1} and \eqref{eq_Lovelockbranch2}. On the other hand, the critical point curve indicates that $\Phi_2$ possesses a unique zero. This is because $\Phi_2$ depends on $\alpha_2$, with $\alpha_0$ eliminated via Eq.~\eqref{eq_lovelockcon2}, and each horizontal line at fixed $\alpha_2$ intersects the curve at a single point, as illustrated in Fig.~\ref{FIG_LoveLock7alpha}. This conclusion can also be directly verified from the explicit form of $\Phi_2$.

	\section{Critical Phenomena and Higher-order Thermodynamical Topology}\label{Sec_CriticalExponents}
	
	In the previous section, we analyze the thermodynamical topology of the 7-dimensional Lovelock black holes and identified the critical values of the coupling constants (see Eq.~\eqref{eq_LoveLockalphac}). When $\alpha_0 < \alpha_{0c}$ while all other parameters are fixed at their critical values, the $F_\text{on-shell}-T_H$ diagram exhibits a characteristic swallowtail structure, reminiscent of the critical phenomena observed in RN-AdS black holes~\cite{Kubiznak:2012PV}. As noted above, at these critical values of the coupling constants, the third order vector field develops a double zero (equivalently, the fourth order vector field has a simple zero), leading to the simultaneous vanishing of the first through fourth derivatives of the temperature with respect to the entropy,
	\begin{equation}
		\partial _ST_H=\partial _{S}^{2}T_H=\,\partial _{S}^{3}T_H=\,\partial _{S}^{4}T_H=0,\quad \partial _{S}^{5}T_H\ne 0.
		\label{eq_LoveLock7critical}
	\end{equation}
	These conditions resemble that of a conventional critical point, but it involves vanishing of higher-order derivatives. We therefore refer to this as a higher-order criticality condition, and to the associated point as a higher-order critical point. This atypical critical behavior motivates a more detailed investigation of the system’s properties in its vicinity. To this end, we fix $\alpha_2 = \alpha_{2c}$ and perturb only $\alpha_0$, which is related to the pressure through the relation $P=15\alpha_0/(8\pi)$~\cite{Kastor:2009Enthalpy,Dolan:2011Pressure}. The extended first law then takes the form
	\begin{equation}
		\delta M = T_H \delta S + V \delta P ,
	\end{equation}
	where the thermodynamic volume is given by $V = \pi ^3r_{h}^{6}/6$. Introducing the near critical expansions
	\begin{equation}
		r_h=r_{hc}\left( 1+\omega \right),\quad P=P_c\left( 1+\epsilon \right),
	\end{equation}
	we obtain the asymptotic form of the law of corresponding states,
	\begin{equation}
		\frac{T_H}{T_{Hc}}=1+\frac{1}{16}\epsilon +\frac{5}{32}\omega \epsilon +\frac{1}{16}\omega ^5+\mathcal{O} \left( \epsilon \omega^2,\omega^6 \right) .
		\label{eq_LoveLock7Ncritical}
	\end{equation}
	where $\omega$ and $\epsilon$ respectively quantify the deviations in horizon radius and pressure at the higher-order critical point. The coexistence curve is determined by the equation of state,
    \begin{equation}
        \left. T_H \right|_{r_h=r_{hc}\left( 1+\omega _s \right)}\left. =T_H \right|_{r_h=r_{hc}\left( 1+\omega _l \right)},
    \end{equation}
    and the Maxwell equal-area law,
	\begin{equation}
		\int_{\omega _s}^{\omega _l}{\left. S \right|_{r_h=r_{hc}\left( 1+\omega \right)}\frac{\partial}{\partial \omega}\left( \left. T_H \right|_{r_h=r_{hc}\left( 1+\omega \right)} \right) d\omega}=0.
	\end{equation}
	where $\omega_s$ and $\omega_l$ denote the values corresponding to the two coexisting black hole phases, respectively, with $\omega_l > \omega_s$ assumed. Substituting Eq.~\eqref{eq_LoveLock7Ncritical} into the above equations, we find a nontrivial solution,
	\begin{equation}
		\omega _l=\left( 5/2 \right) ^{1/4}\left( -\epsilon \right) ^{1/4},\quad \omega _s=-\left( 5/2 \right) ^{1/4}\left( -\epsilon \right) ^{1/4}.
	\end{equation}
	This leads to the following scaling behavior near the critical point,
	\begin{equation}
		\frac{\left| r_{hl}-r_{hs} \right|}{r_{hc}}\sim \left( -\epsilon \right) ^{1/4},\quad \left| \frac{T_H}{T_{Hc}}-1 \right|\sim \left( -\epsilon \right) ,\quad \left| \frac{P}{P_c}-1 \right|\sim \left( -\epsilon \right) .
	\end{equation}
	The thermodynamic analysis yields the following critical scaling relations,
	\begin{gather}
		C_V=T_H\left( \frac{\partial S}{\partial T_H} \right) _V\sim \left| T_H-T_{Hc} \right|^0, \quad \left| r_{hl}-r_{hs} \right|\sim \left| T_H-T_{Hc} \right|^{1/4},
		\\
		 \kappa _T=-\frac{1}{V}\left( \frac{\partial V}{\partial P} \right) _T\sim \left| T_H-T_{Hc} \right|^{-1}, \quad C_P=T_H\left( \frac{\partial S}{\partial T_H} \right) _P\sim \left| T_H-T_{Hc} \right|^{-1}.
	\end{gather}
	Analyzing the scaling behavior of $C_V$, $\left| r_{hl}-r_{hs} \right|$, $\kappa_T$, and the power of the leading $\omega$ term in Eq.~\eqref{eq_LoveLock7Ncritical} with $\epsilon = 0$, we obtain the critical exponents,
	\begin{equation}
		\alpha = 0,\quad  \beta = 1/4,\quad  \gamma = 1,\quad  \delta = 5.
		\label{eq_LoveLockCriticalExp}
	\end{equation}
	Although the higher-order critical condition in Eq.~\eqref{eq_LoveLock7critical} is unconventional, the critical exponents for $C_P$, $C_V$, and $\kappa_T$ retain their values from mean-field theory, whereas $\beta$ and $\delta$ deviate from the standard results of $1/2$ and $3$, respectively. Remarkably, the critical exponents still satisfy the Widom relation,
	\begin{equation}
		\gamma = \beta (\delta - 1).
		\label{eq_WidomR}
	\end{equation}
	However, they yield
	\begin{equation}
		\alpha + 2\beta + \gamma = 3/2 < 2,
	\end{equation}
	thereby violating the Rushbrooke inequality. This violation stems from the higher-order degeneracy at the higher-order critical point imposed by Eq.~\eqref{eq_LoveLock7critical}. To gain deeper insight into this phenomenon, we generalize the discussion within a broader thermodynamic framework.  Specifically, we show that if the $k$-th order vector field $\Phi_k$ admits a simple root for even $k$, then the corresponding critical exponent is given by $\delta = k + 1$.
	
	Let us suppose that $S = S_c$ is a simple root of $k$-th order vector field $\Phi_k$. Then, at this point, the Hawking temperature satisfies
	\begin{equation}
	\partial _ST_H=\partial _{S}^{2}T_H= \ \dots \ =\partial _{S}^{k}T_H=0, \quad \partial _{S}^{k+1}T_H\ne 0.
	\label{eq_CriticalPHigh}
	\end{equation}
	In general, considering the construction of $\Phi_k$ as defined in Eq.~\eqref{eq_Phikdef}, the parameter space is restricted to the constraint surface given by $\left\{ \partial _{S}^{m}F=0,\ m=1,\ \dots,\ k. \right\}$. The zeros of $\Phi_k$, which correspond to higher-order critical points, typically form a hypersurface within the constraint surface. To examine how the system behaves when deviating from such a higher-order critical point, we focus on a specific parameter $X$ with the following properties: when $X = X_c$, the system lies exactly on the locus of higher-order critical points; when $X$ deviates from $X_c$, it moves away from this hypersurface, and we assume that the mixed derivative
	\begin{equation}
		\left( \frac{\partial ^2T}{\partial X\partial S} \right) \ne 0,
		\label{eq_Assuming}
	\end{equation}
	at the higher-order critical point. We then consider a small perturbation of $X$ around $X_c$. Under this consideration, the extended first law of black hole thermodynamics is given by
	\begin{equation}
		\delta M = T_H \delta S + Y \delta X,
	\end{equation}
	where $Y$ is the quantity thermodynamically conjugate to $X$, defined by the partial derivative of $M$ with respect to $X$. Expanding near the critical point as
	\begin{equation}
		S=S_c \left( 1+\omega \right),\quad X=X_c\left( 1+\epsilon \right),
	\end{equation}
	the law of corresponding states can be expressed as follows,
	\begin{equation}
		T_H=T_{Hc}+\left( \frac{\partial T}{\partial X} \right) _cX_c\epsilon +\left( \frac{\partial ^2T}{\partial X\partial S} \right) _cX_cS_c\epsilon \omega +\frac{1}{\left( k+1 \right) !}\left( \frac{\partial ^{k+1}T}{\partial S^{k+1}} \right) _cS_{c}^{k+1}\omega ^{k+1}+\mathcal{O} \left( \epsilon \omega ^2,\omega ^{k+2} \right),
		\label{eq_GenNcritical}
	\end{equation}
	with $\omega$ and $\epsilon$ denoting perturbations of $S$ and $X$ near the critical point, respectively, and the subscript $c$ denotes the evaluation at the higher-order critical point. For $\epsilon=0$, one finds the critical exponent $\delta = k+1$. Along the coexistence curve, the following thermodynamic relations hold,
    \begin{equation}
        \left. T_H \right|_{S=\left( 1+\omega _s \right) S_c}\left. =T_H \right|_{S=\left( 1+\omega _l \right) S_c},\quad \int_{\omega _s}^{\omega _l}{\left( 1+\omega \right) S_c\frac{\partial}{\partial \omega}\left( \left. T_H \right|_{S=\left( 1+\omega \right) S_c} \right) d\omega}=0.
	\end{equation}
	where $\omega_s$ and $\omega_l$ correspond to the two coexisting black hole phases with $\omega_l > \omega_s$. Substituting the expansion~\eqref{eq_GenNcritical} and assuming even $k$, one finds that a nontrivial solution exists with
	\begin{equation}
		\omega_l = - \omega_s \sim \epsilon^{1/k}.
	\end{equation}
	As a result, the near critical scaling behaviors of thermodynamic quantities are
	\begin{gather}
		C_Y=T_H\left( \frac{\partial S}{\partial T_H} \right) _Y\sim \left| T_H-T_{Hc} \right|^0,\quad \left| r_{hl}-r_{hs} \right|\sim \left| T_H-T_{Hc} \right|^{1/k},
		\\
		\kappa _T=-\frac{1}{Y}\left( \frac{\partial Y}{\partial X} \right) _T\sim \left| T_H-T_{Hc} \right|^{-1}, \quad C_X=T_H\left( \frac{\partial S}{\partial T_H} \right) _X\sim \left| T_H-T_{Hc} \right|^{-1}.
	\end{gather}
	From above equaitons, we extract the critical exponents: $\alpha = 0$, $\beta = 1/k$, $\gamma = 1$. When $k = 4$, these reduce to the values obtained for 7-dimensional Lovelock black holes, as shown in Eq.~\eqref{eq_LoveLockCriticalExp}. It is noteworthy that the Widom relation \eqref{eq_WidomR} remains satisfied. However, the combination
	\begin{equation}
		\alpha +2\beta +\gamma =\frac{2}{k}+1,
	\end{equation}
	implies a violation of the Rushbrooke inequality for any $k > 2$, with the standard result recovered only at $k=2$. Notably, the scaling behavior of $C_X$, $C_Y$ and $\kappa_T$ remain unchanged. These results follow from the assumptions \eqref{eq_Assuming} made in the derivation of the near critical expansion~\eqref{eq_GenNcritical}. Importantly, the critical exponent $\delta$ is independent of these assumptions, as it is solely determined by the critical condition given in Eq.~\eqref{eq_CriticalPHigh}. In contrast, the remaining exponents depend on the detailed structure of the equation of state near the critical point.
	
	Therefore, we conclude that for even positive integers $k$, a $k$-th order vector field $\Phi_k$ with a simple root corresponds to a black hole critical point characterized by a critical exponent $\delta = k+1$. This result demonstrates that different values of $k$ for $k$-th order vector fields, along with their associated topological numbers, can potentially characterize distinct types of black hole phase transitions.

	\section{Discussion and Conclusion}\label{Sec_discussion}
	In this work, we reviewed the three thermodynamical topologies associated with black holes, emphasizing their interconnections and motivating a unified topological framework. To this end, we introduced the notion of extended thermodynamical topology, which not only unifies the three existing topologies but also generalizes them. Within this framework, a $k$-th order vector field is introduced. By employing Duan's $\phi$-mapping topological current theory, we characterized the zeros of these vector fields in terms of winding numbers and their associated topological numbers.
	
	A simple zero of $k$-th order vector field $\Phi_k$ at $S = S_c$ characterizes the local thermodynamic stability of the black hole phase near this zero. When $k=0$, each zero corresponds to a black hole phase, and the associated winding number encodes the local thermodynamic stability of that phase. For even $k>0$, both sides of the zero correspond to either stable phases (winding number $+1$) or unstable phases (winding number $-1$); such zeros typically correspond to critical points where the swallowtail structure in the $F_\text{on-shell} - T_H$ diagram terminates. For odd $k$, the stability differs across the zero: a zero with winding number $+1$ indicates a transition from a stable phase at $S < S_c$ to an unstable one at $S > S_c$, while winding number $-1$ implies the reverse; these zeros typically corresponding to cusps in the swallowtail curve. In fact, each simple zero of $\Phi_k$ signals the simultaneous satisfaction of the conditions
	\begin{equation}
		\partial_S F= 0,\quad  \partial_S^{2} F= 0, \quad \dots,\quad \partial_S^{k+1} F= 0,
	\end{equation}
	thereby signaling the coalescence of $k+1$ thermodynamic branches. However, such a coalescence typically manifests as a single multiple zero of zeroth order vector field $\Phi_0$, whose winding number may be $+1$, $0$, or $-1$. Therefore, the zeroth order winding number alone does not reveal how many phases are merging. On the other hand, when $k$ is a positive even integer, a simple zero of $\Phi_k$ implies the condition given in Eq.~\eqref{eq_CriticalPHigh}, which characterizes the nature of the critical point and yields a critical exponent $\delta = k + 1$. Moreover, under the assumption in Eq.~\eqref{eq_Assuming}, we found $\beta = 1/k$, while the other critical exponents remain the same as those in the standard case. This observation reveals a connection between the $k$-th order thermodynamical topology and the associated critical exponents.
	
	In the classification of black holes within general relativity, the extended thermodynamical topological numbers of the RN, Kerr, and KN solutions are found to coincide, despite the KN black holes depending on more parameters. This result is consistent with the fact that KN black holes possess only one locally stable and one unstable phase, and therefore do not exhibit a richer phase structure. In 7-dimensional Lovelock gravity, some special topological structures have been identified. In particular, the zero of the third order vector field $\Phi_{3,2}$ is a double root, which also corresponds to a zero of $\Phi_4$, thereby characterizing a nonstandard critical point. This feature arises from the unique properties of criticality in Lovelock gravity~\cite{Dolan:2014IsoCritical,Frassino:2016CPLoveLock,Ahmed:2022VantiV}. On the other hand, as shown in Fig.~\ref{FIG_LoveLock7alphaL}, when the coupling parameters vary along the vertical line through the green dot, the black hole’s $F_{\text{on-shell}}-T_H$ curve evolves: the swallowtail structure exists initially, vanishes at the green dot, and reappears for parameters beyond this point. A similar phenomenon occurs in hyperbolic black hole in Lovelock gravity~\cite{Dolan:2014IsoCritical,Frassino:2016CPLoveLock,Ahmed:2022VantiV} and in black hole thermodynamics with quantum anomalies~\cite{Cai:2009ConformalAnomaly,Cai:2014ConformalAnomaly,Hu:2024QuantumAnomalyTriggers}.
	
	In conclusion, the extended thermodynamical topology provides a unified and systematic generalization of the three known thermodynamic topologies, establishing a coherent framework for the study of black hole thermodynamics. This approach enables the systematic classification of black hole phase structures, identifies key thermodynamic features such as critical points and cusps, and uncovers profound connections between topological structures and critical exponents. As such, it serves as a powerful tool for revealing the underlying thermodynamic landscape of complex black hole systems, while offering new theoretical perspectives for advancing the understanding of gravitational phase transitions.

	\section*{Acknowledgments}
	This work was supported by the National Natural Science Foundation of China (Grants No. 12475055, No. 12305065, and No. 12247101), the China Postdoctoral Science Foundation (Grant No. 2023M731468), the Basic Research Foundation of Central Universities (Grant No. lzujbky-2024-jdzx06), the Natural Science Foundation of Gansu Province (Grant No. 22JR5RA389), the ‘111 Center’ under Grant No. B20063.

	\newpage
	
	%\bibliographystyle{unsrt}
	%\bibliography{ref}

\end{document}